\documentclass[12pt]{aa}
\usepackage[english]{babel}
\usepackage{graphicx}
\usepackage{longtable}

\onecolumn
\newcommand{\ab}{Astrophys. Bull. }
\newcommand{\bsao}{Bull. Spec. Astrophys.Observ.}
\newcommand{\arep}{Astron. Rep. }
\newcommand{\alet}{Astron. Let. }
\newcommand{\araa}{Ann. Rev. Astron. Astrophys. }
\newcommand{\mnras}{Mon. Not. R. Astron. Soc. }
\newcommand{\apj}{Astrophys. J. }
\newcommand{\apjl}{Astrophys. J. Lett.}
\newcommand{\apjs}{Astrophys. J. Suppl.}
\newcommand{\aj}{Astron. J.}
\newcommand{\aaa}{Astron and Astrophys.}
\newcommand{\aas}{Astron and Astrophys. Suppl.}
\newcommand{\pasp}{Publ. Astron. Soc. Pasif. }

\begin{document}

\title{Spectroscopy of supergiants with infrared excess: results of 1998--2018} 
\author{V.G.~Klochkova}
\institute{Special Astrophysical Observatory RAS, Nizhnij Arkhyz,  369167 Russia}
\date{\today} 

\abstract{The results of our second stage (1998--2018) of the detailed spectroscopy 
of peculiar supergiants identified with galactic infrared sources, performed
at the 6-meter  BTA telescope are summarized. The main aspect of the program is a
search for the evolutionary variations in the chemical composition of stars, 
past the AGB stage and the TDU, as well as an analysis of spectral manifestations 
of kinematic processes in their extended, often unstable, atmospheres and 
in the envelopes. The most significant result is detection of the $s$-process element
excesses in seven single post-AGB stars, which  confirms the theory of evolution   
of this type of stars. In three of these stars we for the first time discovered 
the ejection of the $s$-process heavy metals to the circumstellar envelopes. A
Li excess was found in the atmospheres of peculiar supergiants V2324\,Cyg and V4334\,Sgr. 
The results of investigation of the kinematical parameters of atmospheres and envelopes 
will clarify the equilibrium of matter produced by stars in the AGB and
post-AGB stages and delivered to the interstellar medium.
}
\titlerunning{Spectroscopy of supergiants with infrared excess  in 1998--2018}
\authorrunning{Klochkova}

\maketitle

\section{INTRODUCTION}

Peculiar supergiants we study possess a large infrared flux excess and  primarily 
belong to the evolutionary stage after the asymptotic giant branch, further referred 
to as post-AGB. Our results of research of a sample of supergiants obtained 
at the first stage of spectroscopic observations at the 6-meter telescope are presented 
in the survey~[1]. Here we complement this survey, adding the results obtained and published 
over 1998--2018.  We shall briefly recall the main features of post-AGB stars, also called 
the protoplanetary  nebulae (PPN), then give a rationale for the relevance of a detailed study 
of these objects and describe the main results obtained by the program in the previous decade.

The PPN stage hosts far evolved stars of intermediate masses,
with initial masses on the Main Sequence (MS further in the text) of less 
than 8$\div9 \mathcal{M}_\odot$.  At the previous evolution stage,
the asymptotic giant branch (AGB), these stars are observed as cool 
(their effective temperature is T$_{\rm eff}\approx3000$~K)  supergiants.
The AGB stage, being significantly shorter than the previous stages
of evolution is very important due to a large mass loss and the formation 
of a circumstellar shell~[2]. Also note  that in this very mass range, near 
the MS, about 10\% of stars possess large magnetic fields.
Using the method of measuring circular polarization  of maser lines,
magnetic fields were also detected  in a number of their far evolved  descendants, 
PPN and PN~[3, 4].

After the helium depletion in the core of an AGB star, a
degenerate C--O core surrounded by alternative energetically active
shells of burning helium and hydrogen is formed. Most of the time
the energy release is provided by the hydrogen burning, while the
helium shell  adjacent to the degenerate C--O core  remains
inert (for details and references, see the survey~[5]). Between these major
energy-release layers  a thin shell  is formed, the so called ``He
intershell'', in which, as it gets compressed and as the
temperature rises, the helium ignites and carbon accumulates,
which creates conditions for the reaction of prime importance,
$^{13}$C($\alpha$, n)$^{16}$O. This reaction is the main source of
neutrons providing the subsequent synthesis of the heavy metal
nuclei (a detailed description of these processes and the
necessary references are available in the papers~[6, 7]). 
An additional energy release in the ``He intershell'' changes the
configuration of the internal structure of the star and leads to
an instability with a  development of efficient convective mixing
of matter in the layers. Due to the penetrating convection, the
stellar matter, recycled in nuclear fusion accompanying these
processes of energy release~[8], outflows into the atmosphere of the star. 
It is common practice to call the mixing of matter due to the repeatedly 
recurring alternation of thin, energetically active layers of burning 
hydrogen and helium ``the third dredge-up'' (hereinafter -- TDU).

The primary gas in the early Universe was a mixture of hydrogen
and helium admixed with light elements (Li, Be, B). Later, in the
course of evolution and explosions of massive first-generation
stars there began formation of nuclei of heavier chemical elements
(C, O, Ne, Mg, Si and Fe). The entire variety of chemical
abundances that we observe now is due to nuclear fusion 
results in the evolution of several stellar populations. 
An increased interest in AGB stars is due to the fact that the 
inside of these stars that are on the short-term
(characteristic time of $10^3$~years) evolutionary stage, form
ideal physical conditions for the synthesis of heavy metal nuclei
and for the transport of accumulated nuclear reaction products
into the stellar atmospheres. The calculations show that AGB
stars are the main suppliers of heavy metals (over 50\% of all
elements are heavier than iron) into the interstellar medium,
while the most efficient suppliers are   AGB stars  with
initial masses not exceeding 3\,$\mathcal{M}_\odot$.  The synthesis of heavy
metal nuclei is carried out due to the so-called $s$-process, the essence of 
which is in a slow (compared with the $\beta$-decay) neutronization of nuclei~[9]. 

The seed nuclei for a chain of $s$-process reactions are the  Fe nuclei. 
In stars with the initial masses below 3\,$\mathcal{M}_\odot$ the necessary 
neutron flux is provided by the $^{13}$C($\alpha,\,n)^{16}$O reaction, and in the
case of more massive stars with initial masses greater than
$4\div5\mathcal{M}_\odot$, a similar reaction takes course in the $^{22}$Ne
nuclei. These more massive AGB stars may also be sources of Li. 
A description of the evolution of stars near the AGB and
the results of modern calculations of the synthesis and outflow of
elements are given in~[10, 11].

The matter of an AGB star is lost through two processes: firstly,
carbon and oxygen produced during the nucleosynthesis join the
degenerate C--O core (which changes the character of energy
release), secondly, the matter is lost owing to  wind from the
surface of the star. As a result, the AGB star loses from 50\% to
85\% of its initial mass~[12]. Due to the matter loss, an optically 
thick envelope is  formed around the star, and the star becomes 
difficult to observe in the optical range. At some point in the    
AGB star lifetime its   mass loss
rate suddenly increases dramatically. One of the reasons for the
large mass loss rate increase can be the pulsating instability of
supergiants, which is  inherent, according to~[12], 
to the stars on this evolutionary stage  in a fairly wide range of 
fundamental parameters. Two physical phenomena: the synthesis of 
heavy metals and the
pulsation activity characteristic of stars on the AGB stage and
after it, determine the main interest in these objects.

The complexity of the theoretical study of stellar nucleosynthesis
is caused by the fact that the efficiency of synthesis and the
ejection of fresh nuclei onto the stellar surface depends on a
large number of factors: the original mass of a star on the MS,
its metallicity, the parameters of the matter outflow, the
features of the nucleosynthesis process and uncertainties of
numerical description of the mixing process. Therefore, besides
the classical problem of studying the chemical composition of
stars on nontrivial stages of evolution, of particular interest is
also the possibility of studying the processes of exchange of
matter between the stellar atmosphere and the circumstellar dust
envelope, and also the search for mechanisms explaining the
peculiarities of  chemical composition of the atmospheres of stars
surrounded by dust envelopes. It is generally believed that the
conditions of a gas and dust envelope  can be conducive to quite
effective processes of selective depletion of metal nuclei onto
the dust grains. In the course of formation of dust particles, the
elements with a high temperature condensation predominantly depart
from the gas component. In particular, this selective process
distorts the pattern of the TDU, increasing the C/O, C/Fe, N/Fe,
O/Fe ratios and decreasing the  $s$/Fe ratio. The main argument
confirming the operation of the mechanism of selective depletion
of metal nuclei on the dust particles is a dependence of the
chemical element abundances on the  temperature of condensation on
the dust particles~[13].

The abundance of Fe, Mg, Si, Ca in the atmospheres of such stars
is reduced by several orders of magnitude, while the CNO, S, and
even the iron group element Zn have solar abundances. It is worth
emphasizing that this pattern is similar to the behavior of the
chemical element abundances in the gas component of the ISM. The
stellar atmosphere should obviously be stable enough so that the
mixing or the stellar wind would not distort the chemical
abundance pattern. However, there are evidences that the
atmospheres of post-AGB stars are unstable: most of these objects
manifest pulsations, the outflow of matter, which is manifested in
the presence of a variable emission, mainly in the lines of
hydrogen. The selective depletion process efficiency, its
dependence on the effective temperature and metallicity of the
star are illustrated in the data of Rao et al.~[14].

A theoretical study of the evolution of stars near the AGB is
difficult due to the complexity of this system including at least
two components: the  central star continuing to evolve and a
cooling envelope that may have   quite a complex structure. A lot
of ambiguity remains in the understanding of stellar wind,
convection processes, and, therefore, the outflow of matter.
However, in general, the idea has come up that along with the
star's initial mass, the rate of mass loss during its evolution is
the most important parameter which does  not only  determine the
final mass after the AGB phase, but also the internal structure of
the star and the characteristic times of evolution in the final
phases, as well as the changes in the chemical composition of the
stellar atmosphere.

The Infrared Astronomical Satellite, or the IRAS, performed a survey of 
about 95\% of the entire sky. One of the results of this IR survey was identification at 
high Galactic latitudes of the sources   representing the circumstellar envelopes 
with temperatures from 200 to 1000~K.
Subsequently, some of these objects were identified with high luminosity stars
presumably at the  AGB and post-AGB evolutionary stages.
A small part of   selected objects is available for the high spectral resolution 
spectroscopic observations in the optical range, providing an empirical base in the
stellar nucleosynthesis studies. After the new type of
far evolved stars with a large excess of infrared flux was identified,
programs of the spectroscopy of post-AGB and AGB stage candidate stars
were initiated at several large telescopes of the world. Their
purpose is to investigate the features of anomalous spectra, the kinematic
state of the outflowing atmospheres and extended circumstellar envelopes
and, above all, to study the process of stellar nucleosynthesis
and  release of its products into the stellar atmosphere.

\section{SPECTROSCOPY OF POST-AGB CANDIDATES:  RESULTS OF 1998--2018}

This kind of an observational program under the direction of the
author of the present paper is also performed at the 6-m BTA
telescope coupled with modern echelle spectrographs. In the course
of longterm observations (spectroscopy of each of the objects was
carried out repeatedly) we have obtained and published several
fundamentally new results. The chemical composition was studied in
about twenty supergiants with an IR-excess, located mainly in the
galactic field. For more details, see the survey by
Klochkova~[1].
Since 1998,  the observations are carried out with the NES echelle
spectrograph~[15, 16]  in combination with the CCD chip sized $1{\rm K}\times1{\rm
K}$, since 2000 --$2{\rm K}\times2{\rm K}$, and since
2011 -- $2{\rm K}\times4{\rm K}$. The result is a collection
of high-quality spectra, which were intended primarily to search
for the chemical composition anomalies due to the nuclear fusion
of chemical elements in the depths of stars of small and
intermediate masses and a subsequent outflow of the synthesis
products into the stellar atmospheres. This observational material
is also used to search for the peculiarities of spectra of
post-AGB candidates, to analyze the velocity field in the
atmospheres and envelopes of stars  with mass outflow, as well to
search for the expected long-term variability of spectral
features. To reduce the losses at the entrance slit of the
spectrograph, the observations are performed with an image slicer,
so every echelle order is repeated thrice.


In general, based on the studied sample of PPN candidates using
the published results of other authors, we can talk about the
heterogeneity of the chemical composition of the sample. An excess
of the $s$-process elements expected for stars in the post-AGB
stage as a consequence of the previous evolution of the star and
the process of the TDU, is  so far  extremely rarely found. For
example, we have revealed an excess of carbon and $s$-process
elements in the atmosphere of the first post-AGB star we chose to
study, CY\,CMi (IRAS\,07134+1005)~[17]. By
now, a significant excess of  $s$-process elements was found only
for eleven post-AGB stars, seven of which were examined from the
BTA spectra: IRAS\,04296+3429~[18],
IRAS\,07134+1005~[17], IRAS\,20000+3239~[19], IRAS\,22272+5435~[20],
IRAS\,23304+6147~[21], RAFGL2688~[22], as well as a virginide
V1\,K307~[23]  in the globular cluster M\,12. In addition, conclusions about the excess of heavy metals are
published for the following PPN candidates: HD\,158616~[24], HD\,187885~[24],
IRAS\,22223+4327~[25], IRAS\,02229+6208~[26]  and IRAS\,05341+0852~[25]. In the spectrum of the
nameless and faint in the optical range   post-AGB star with an
emission at 21\,$\mu$, identified with an IR source
IRAS\,20000+3239, an overabundance of heavy elements was
detected~[19]. For example, the concentration of La nuclei is increased by an order of
magnitude with respect to the stellar metallicity. In this regard,
the La\,II line is so strong that its intensity is comparable to
the intensity of the H$\alpha$ line.

Let us emphasize that all of these stars, selected based on their
chemical composition and the presence of a band at 21\,$\mu$ are
single objects, as confirmed by the authors of~[27]. Currently, 27 
stars are known with an emission at 21\,$\mu$, and they all belong  to the
post-AGB stage only~[28]. The presence of a
feature at 21\,$\mu$ in the spectra of post-AGB stars with
carbon-rich shells, allows to suggest some kind of a complex
molecule containing the carbon atoms as the main agent. However,
other, but not confirmed, alternatives of identifying the features
at 21\,$\mu$ were proposed. For example, nanoparticles of iron
oxide FeO~[29], capable of, when heated, producing radiation at the wavelength 
of 21\,$\mu$, and titanium carbide TiC~[30]  were considered as agents.

Based on the first facts of detection of heavy metals
overabundance in the papersof Klochkova~[1], Decin et al.~[31] 
made a conclusion about the relationship 
of emissions at 21\,$\mu$ in the IR spectra of stars at the
post-AGB stage and manifestations of heavy metals excess in their
atmospheres. Our results obtained for AFGL\,2688 strengthened this
conclusion, since this object is characterized by both   a slight
excess of $s$-process elements, and a 21\,$\mu$ band, practically
not revealed on the background of a monotonously increasing IR
continuum. So far, this result remains one of the most significant
(but not explained)  in the pattern of stellar nucleosynthesis
manifestations at the AGB and post-AGB stages. Van~Winckel
and Reyniers~[24]  reiterated a study of all six
stars  with a feature at 21\,$\mu$ known by that time, and on a
uniform spectroscopic material confirmed a conclusion about the
effectiveness of the $s$-process for this group of objects.
Specifically, these authors have obtained a strict correlation
between the magnitude of neutron exposure estimated from the
observations as an [h$s$/l$s$]  ratio (an excess of the
$s$-process heavy nuclei relative to the lighter ones), and
[$s$/Fe], as well as a less pronounced anti-correlation between
[h$s$/l$s$] and metallicity  [Fe/H].

As a rule, in the atmospheres of PPN candidates there is an
overdeficiency (relative to their metallicity) of heavy nuclei,
the existence of which in the atmospheres of low-mass supergiants
at the post-AGB stage finds no definite explanation as yet. There
is a number of physical (a hydrogen deficiency  in the
atmospheres; overionization of atoms with a low second ionization
potential) and methodical (parameter errors) effects that could
explain the observed overdeficiency of the $s$-process elements in
the atmospheres of low mass supergiants. None of the suggested
explanations is consistent with the full chemical abundances
picture of these objects. A low occurrence of this type of
atmosphere enrichment is a consequence of differences in the
evolution of AGB stars, that vary in their initial masses,
since it is the initial mass of the star that affects the possible
manifestations of nucleosynthesis and the matter loss rate. It is
currently established that the surface layers of the less massive
AGB stars (with initial masses below $4\,\mathcal{M}_{\odot}$) are
enriched with carbon and heavy metals---this is how  C-rich stars
appear, while more massive stars at the subsequent stages of their
evolution  remain  O-rich stars.

\subsection{Spectral mimicry of supergiants}

An important property that determines the reliability of results
of studying the AGB- and post-AGB stars, is a high quality of the
high-resolution spectra obtained on the echelle spectrographs of
the BTA. For the majority of stars of the program the data of such
quality has been obtained for the first time, which allowed us to
make a number of new conclusions regarding the evolutionary status
of individual objects, spectral features, the velocity field and
chemical composition of the atmospheres. First of all, note our
conclusion about the heterogeneity of the studied sample of
objects in general. As we collected the spectra and the results,
it became clear that the original sample of stars with a large
excess of IR flux, in addition to AGB and post-AGB stars, includes
massive high luminosity stars with extended and structured
circumstellar envelopes. Therefore, the necessary and the most
time consuming point of our research is determination of
luminosity and mass of a star, and therefore recording of its
evolutional stage. When studying individual stars, we repeatedly
found a similarity and even a complete reiteration of spectral
features in objects of different nature: LBV stars, stars with a
B[e] phenomenon that are likely to be binary intermediate-mass
systems soon after the stage of a rapid matter exchange; white and
yellow hypergiants and low mass supergiants at the post-AGB stage
with a large excess of infrared flux.

Paradoxically, the high luminosity stars, fundamentally differing
in mass (the most massive stars with initial masses exceeding
$20\div40~\mathcal{M}_\odot$ and intermediate mass stars
3--9~$\mathcal{M}_\odot$) and evolutional stage, have close observed
properties: the features of the optical and radio spectra, a large
excess of IR flux, a complex and time-varying velocity field,
which testifies to the instability of extended atmospheres and
expanding gas and dust envelopes. This  similarity of the observed
properties of two types of objects predetermines the problem of
spectral mimicry~[32].

\begin{figure}[ht!]
\includegraphics[scale=0.6, bb=40 60 510 530,clip]{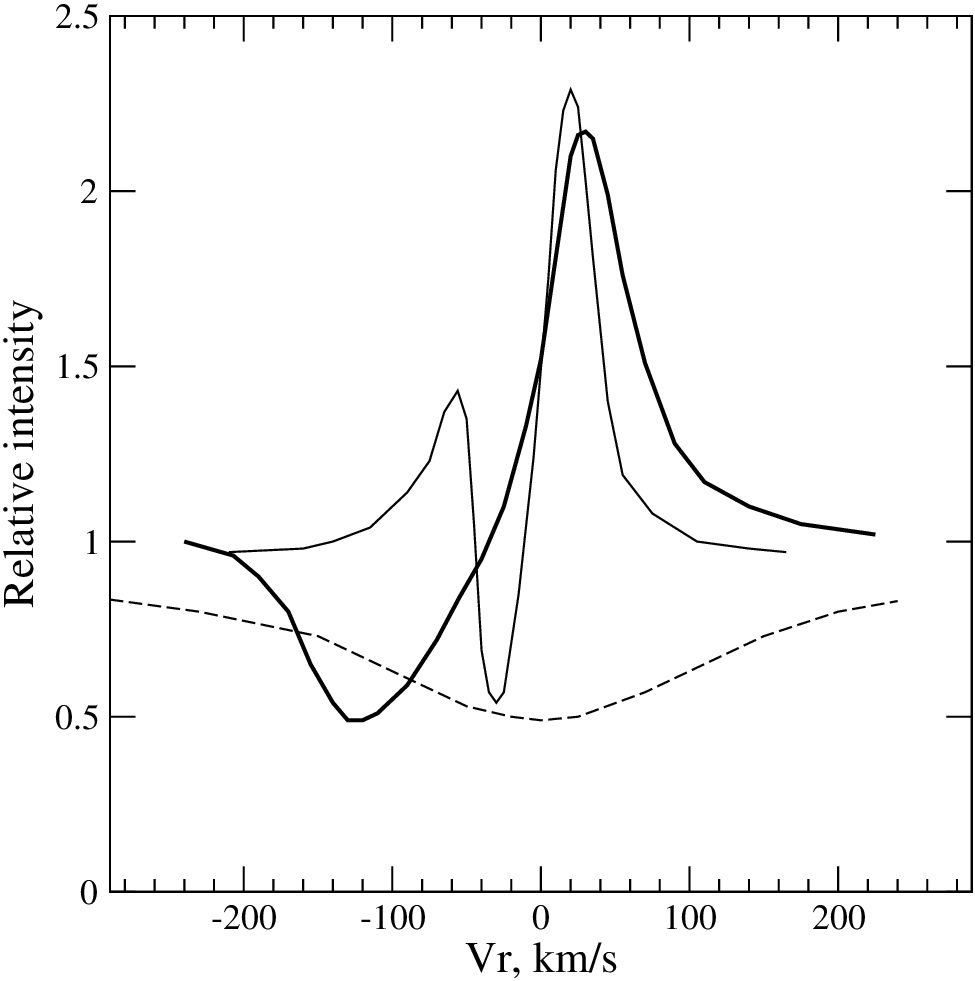}
\caption{H$\alpha$  profiles in the spectra of selected supergiants of different initial 
  mass: a classic supergiant $\alpha$\,Per --the dashed line, a hypergiant HD\,33579 --
  the thin solid line, a post-AGB supergiant V510\,Pup --the continuous solid
   line. The abscissa axis describes  the $\Delta V_r$ shift relative to the systemic 
   velocity of each object.}
\label{Mimicry_Halpha}
\end{figure}

Distinctive features of the optical spectrum of a hypergiant are
considered to be a powerful emission (often possessing two peaks)
in the lines of the Balmer and Paschen series of hydrogen, having
broad wings due to  Thomson scattering and the presence of
forbidden emissions of metals. However, it must be borne in mind
that a combination of these features is also present in the
spectra of other types of high luminosity stars (see
Fig.~\ref{Mimicry_Halpha}). To get acquainted with the
properties of these massive hot supergiants you can use the
papers~[33-35] having good examples of optical spectra obtained with high
spectral resolution. A characteristic feature of the spectra of
B[e] stars are the emissions in the lines of an IR triplet Ca\,II
and in the forbidden lines of Ca[II]\,7291 and 7324~\AA{}. These
profiles are in a good agreement with the hypothesis of the
presence of rotating circumstellar disks in the systems of B[e]
stars. For a more detailed study of the spectra of B[e] stars, the
atlas of spectra of a B[e] star MWC\,314 and a hypergiant
V1302\,Aql~[36]  may also be useful.

It should be borne in mind that only a combination of an intense
(often two-peak) H\,I emission  with   forbidden emissions of
metals does not give  grounds for classifying a star as a
hypergiant. The rationale for the high luminosity and a large
matter loss rate at a slow velocity of its outflow is principal
here. As important is a detailed analysis of the radial velocity
pattern measured from the lines of different nature. A good
illustration of this problem is given by the already mentioned
above spectral atlas~[35],   which compares the spectra of the star 
MWC\,314 (Sp\,=\,B3\,Ibe) with a B[e] phenomenon and the spectra of a hypergiant V1302\,Aql (its
Sp\,=\,F5\,Ia$^{+}$e at the times of observations used in the
atlas). These stars with different masses, luminosities and
spectral classes have similar spectra. In particular, powerful
hydrogen emission lines have two-peak profiles. The spectra
contain plenty of permitted and forbidden metal emission lines and
features with the P\,Cyg-type profiles.

The most famous object, which was for many years  considered as a
probable star in the post-AGB stage is V1302\,Aql, having numerous
spectral peculiarities. For a long time the evolutionary status of
the supergiant V1302\,Aql, associated with a powerful source of IR
radiation IRC+10420, was unclear. A manifold of observable
properties allowed to consider it as a star at the post-AGB 
stage or as a very massive star, past the stage of the red
supergiant. Obviously, depending on the adopted status, and hence
the luminosity of the object, the estimate of its distance from
the observer may vary by several times. However, the information
obtained over the last decades through various observational
experiments as well as the NES~BTA monitoring results~[37-39] 
leave no doubt as to the fact that V1302\,Aql belongs to extremely
rarely observed objects, yellow hypergiants. Moreover, V1302\,Aql,
whose luminosity is $L\approx 5\times10^5~L_{\odot}$,  is
now considered as the most undisputed massive object in the Galaxy
(with an initial mass of  $20\div40~\mathcal{M}_{\odot}$) 
with a record mass loss rate, undergoing a short evolutionary transition
from the massive red supergiants to the Wolf--Rayet stars~[40].

\subsection{A peculiar supergiant  V510\,Pup}

Another program object,  a supergiant V510\,Pup is a clear example of spectral mimicry.
The luminosity of this star, associated with an IR-source IRAS\,08005$-$2356
barely reaches the luminosity level of a normal supergiant, while its wind parameters 
are close to the hypergiant wind parameters  and even those of an LBV~[41].
In the spectrum of V510\,Pup  the absorptions of Y\,II and other $s$-process elements are  abnormally strong.
All the absorption components are shifted to the shortwave spectral region, what
indicates the outflow of stellar matter.
The circumstellar envelope of the supergiant manifests itself in the optical spectrum, in particular, 
in the form of molecular bands of the carbon-containing molecules
C$_2$ and CN. The emission in H$\alpha$ by manifold exceeds the continuum level, what
testifies to the powerful stellar wind. The $V_r$ differences
within one spectrum reach 100~km\,s$^{-1}$ mainly due to the
differential shift of emission and absorption components, but also due to the systematic
velocity variations with the intensity and line wavelength.
Such large differential line shifts clearly indicate the velocity gradient in
the layers of the atmosphere of V510\,Pup, where these lines are formed, and
many P\,Cyg-type profiles are indicative of the expansion of the atmosphere.

\subsection{A variable star V2324\,Cyg}

For the first time, high spectral resolution optical spectroscopy
of a variable star V2324\,Cyg (Sp\,=\,F3\,Ie) associated with an
IR source  IRAS\,20572+4919 was performed at the BTA
coupled with the NES spectrograph. The results are published by
Klochkova et al.~[42]. Over 200 absorptions were identified 
(mainly Fe\,II, Ti\,II, Cr\,II, Y\,II, Ba\,II) within the wavelength range 
from 4549 to 7880~\AA{}. The spectral class F0\,III and   rotation velocity 
$V\sin i=69$~km\,s$^{-1}$ were determined. The lines of neutral hydrogen
and the Na\,I D-lines have a complex P\,Cyg-type profile. Neither
the systematic course of radial velocity with the depth of the
photospheric line, nor the temporal variability of V$_r$ were
detected. The average value of the heliocentric radial velocity is
V$_r=-16.8\pm0.6$~km\,s$^{-1}$. Velocities determined from
the cores of absorption components of the   H$\beta$ and Na\,I
wind lines range from $-140$ to $-225$~km\,s$^{-1}$ (the
velocities of expansion of the corresponding layers amount from
about 120 to 210~km\,s$^{-1}$). The maximum expansion velocity
obtained from the blue component of the split  H$\alpha$
absorption is V$_r=450$~km\,s$^{-1}$ as of December 12, 1995.
Using the model atmosphere method the following parameters were
determined: effective temperature T$_{\rm eff}=7500$~K,
surface gravity $\log g=2.0$, microturbulence velocity
V$_t=6.0$~km\,s$^{-1}$ and a metallicity close to solar.

The main feature of the chemical composition of the star is a
lithium and sodium excess. An excess of lithium was also found in
the spectra of a post-AGB star associated with the source
IRAS\,04296+3429~[18]. It should be noted
that the Li\,I\,6707~\AA{} line has a close neighbor line
Ce\,II\,6708.1~\AA{}. Therefore, it can be blended in the spectra
of stars with a large excess of the $s$-process heavy metals. In
particular, the authors of~[43]  explained the lithium excess, discovered in~[44]  in
the atmosphere of the central star of IRAS\,05341+0852 by this
blending.

Based on the results obtained, the belonging of V2324\,Cyg to
post-AGB stars is being questioned. The low luminosity of the star
is not consistent with the post-AGB stage: the spectral
classification indicates the luminosity class III. The H$\alpha$
line profile and very high wind velocities, usually inherent to
supergiants are also not combined with the status of a
post-AGB star. The aforementioned lithium excess in the
atmosphere of the low luminosity star suggests an alternative
evolutionary phase for V2324\,Cyg. The phenomenon of lithium
overabundance in the atmospheres of F-giants with an IR excess is
already known. A part of these Li-rich giants who have not yet
reached the AGB stage and are observed  on the red giant branch,
has a low mass, not exceeding 2.5$\mathcal{M}_{\odot}$. The so-called 11cool
bottom process'' is suggested as a lithium production mechanism for
such giants, it is based on the synthesis of beryllium, a transfer
of its nuclei to the base of the convective shell and a subsequent
destruction to the lithium nuclei in the near-surface layers of
the star. It will be appropriate here to refer to thepaper~[45], 
the authors of which identified the Li abundance for a sample of 
G--K giants and subgiants with solar metallicity, and where the largest 
values of Li abundance are obtained for the fast-rotating stars 
V$\sin i>30$~km\,s$^{-1}$.

The belonging of V2324\,Cyg to giants is consistent with the low
luminosity of the star M$_V\approx0\fm1$, obtained using
its parallax of $\pi=1.55$~mas according to the Gaia\,DR2
data and the color excess E(B-V)=$0\fm79$, according to~[46]. 
However,  the spectral features of V2324\,Cyg in the form of strong H\,I  
emissions do not allow to finally accept for V2324\,Cyg the status of a red giant. 
The H$\alpha$ profile has a P\,Cyg-type, its emission peak is
3-4~times higher than the continuum level~[42]. The H$\alpha$ profile mapping
in the spectra taken at different times indicates their
variability: the shape and intensity of the emission  and the
depth of the absorption component vary. The position of the
emission peak and the absorption also varies. The spectral
features rather indicate a high luminosity, and therefore a
greater distance to V2324\,Cyg. Note that the authors
of~[46], via modeling the energy distribution, calculated the total radiation flux, 
having at that obtained a high luminosity $\log L/L_{\odot}=3.78$ and distance
$d=4.4$~kpc. To clarify the nature of this star we further require
accumulation of more data not only about the object itself, but
also about the stars in the direction of the Cas\,OB7 association,
a member of which V2324\,Cyg apparently is.

\subsection{Detection of a new phenomenon in the spectra of post-AGB stars}

\begin{figure*}[ht!]
\includegraphics[scale=0.6, bb=40 40 710 530,clip ]{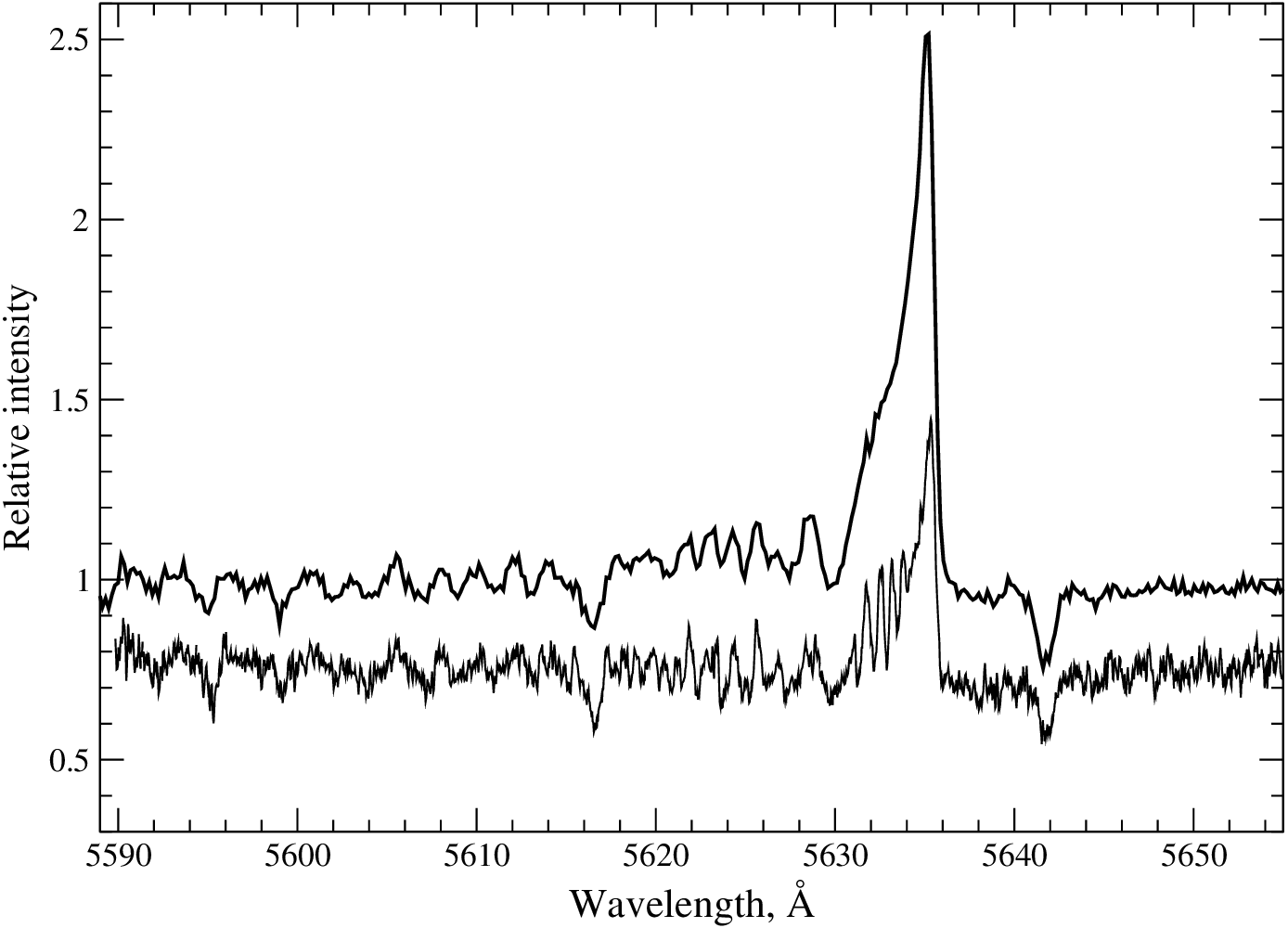}
\caption{Fragments of the RAFGL\,2688 nebula spectra  with the Swan system emission 
band 5635.5~(0;1) of the ${\rm C_2}$ molecule at different dates: {\it 1}---according 
to observations with the PFES spectrograph, {\it 2}---shifted down by 0.2, according to the data obtained
with the NES spectrograph.} 
\label{Egg_Swan}
\end{figure*}

As we emphasized above, an important result of the spectroscopic
research of post-AGB stars is that based on the sample
studied so far, we can talk about its chemical composition
heterogeneity. An excess of the $s$-process elements expected for
stars in the post-AGB stage as a consequence of the previous
evolution of the star and the process of the TDU, is extremely
rare. Over two decades of the search, an excess of the $s$-process
elements has only been detected in seven post-AGB stars (about
fifty objects were studied in details). An analysis of the
totality of observed properties of the post-AGB objects, in the
atmospheres of whose central stars large excesses of carbon and
heavy metals are revealed, led to the conclusion that the
circumstellar envelopes of these selected stars have a complex
morphology~[47]. In addition, their envelopes are also enriched with carbon, 
which is manifested in the IR, radio and optical spectra, in the presence of molecular
lines and the C$_2$, C$_3$, CN, CO bands. Figure~\ref{Egg_Swan} shows the Swan system band of
the C$_2$ molecule in the spectra of the RAFGL\,2688 nebula~[22]  according to 
the observations at different nights at the 6-m telescope 
with the PFES spectrograph~[48]  and with the NES spectrograph~[15],  
providing the spectral resolution of $R=15\,000$ and  $R=60\,000$, respectively.

The collection of spectra obtained via the long-term spectroscopy
at the BTA with high spectral resolution served as the basis for
the study of the variability of spectra and the variability of the
velocity field in the atmosphere and envelope of the star. This
variability is explained by the outflow of extended atmospheres in
combination with the pulsations of deeper atmospheric layers. A
detailed study~[49]  of the optical spectra of a post-AGB star  
HD\,56126 (IRAS\,07134+1005) in the wavelength
region from 4012 to 8790~\AA{}, obtained with the echelle
spectrographs of the 6-m telescope, led to a detection of a
complex and variable profile shape of strong lines (H\,I and the
Fe\,II, Y\,II, Ba\,II, etc. absorptions), forming in the expanding
atmosphere (in the wind base) of the star. The authors
of~[49]  have shown that for studying the kinematics of the atmosphere 
one has to measure the velocities from the individual features of 
these profiles. Based on the metal lines and molecular features, differential 
shifts of lines were found, reaching $\Delta V_r=15$--$30$~km\,s$^{-1}$. The atmosphere
of the star simultaneously manifests both the expanding layers and
layers infalling on the star. A comparison of data for different
times of observations leads to the conclusion on the variability
of   radial velocity and the velocity pattern in general. The
position of the molecular spectrum is temporally stable, which
indicates the stability of the process of  expansion of the
circumstellar envelope in HD\,56126, observed in the C$_2$ and
Na\,I lines.

\begin{figure}[t!]
\includegraphics[scale=0.6,  bb=40 50 460 530,clip]{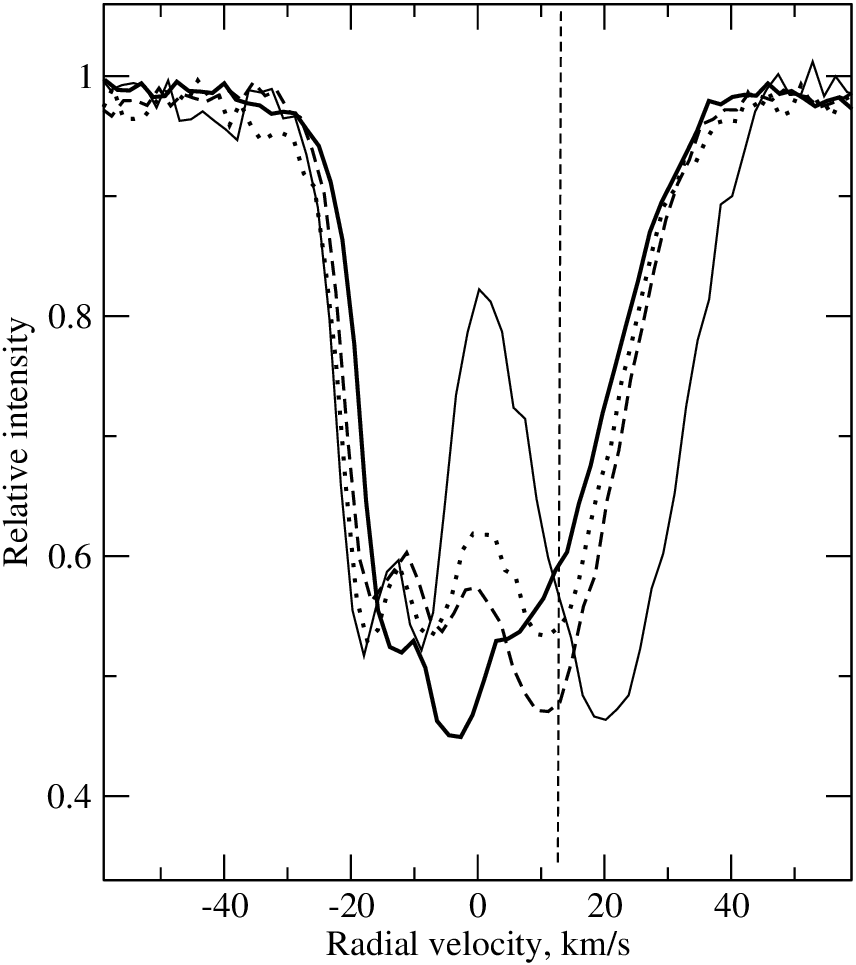}
\caption{The Ba\,II\,4934~\AA{} line profile in the spectra of V5112\,Sgr,
obtained on different dates: the thin solid line --August 2, 2012,
the thick solid line -- June 13, 2011, the dotted line -- August 14, 2006,
the dashed line -- July 7, 2001. The vertical dashed line describes the systemic velocity
V$_{\rm sys}=13$~km\,s$^{-1}$, according to~[50]. }
\label{V5112Sgr_Ba4934}
\end{figure}

For several selected objects of the program we found a phenomenon,
previously unknown for the spectra of post-AGB stars, namely, an
asymmetry and/or splitting of the strongest absorptions with a low
excitation potential of the lower level. Let us take a closer look
at this effect on the example of a spectrum of a post-AGB
{\bf supergiant V5112\,Sgr}, studied by Klochkova~[50]. V5112\,Sgr 
has an envelope enriched by the TDU, an emission at the wavelength of 21\,$\mu$,
and a powerful and structured IR envelope. The discovered
absorption splitting effect  is maximal for the Ba\,II ion lines,
in the profiles of which in the case of V5112\,Sgr  three
components are observed (see, for example, Fig.~\ref{V5112Sgr_Ba4934}). 
At that, the shape of the profiles of the split lines and the position of the atmospheric
(``stellar'') component changes over time, while the position of
circumstellar components is almost stable. The analysis of the
velocity pattern, performed using the radio spectroscopy data of
the object, allowed to make a conclusion that the shortwave
components of the split  Ba\,II absorptions are formed in the
structured circumstellar envelope expanding at  velocities of
$V_{\rm exp}\approx20$ and $30$~km\,s$^{-1}$. Detection of the
envelope components of heavy metals indicates an effective
transport into the envelope of the matter synthesized in the
previous stage of stellar evolution. Thus, an enrichment of a
shell of a star with heavy metals synthesized at the AGB stage was
discovered for the first time.

A good example is also the object {\bf IRAS\,23304+6147}. In the
atmosphere of the faint in the optical range central star of this
source previously revealed an excess of carbon and heavy
metals~[21]. According to the observations with high spatial resolution 
at the Hubble Space Telescope~[51], the circumstellar envelope
in this system has a complex structure including a multipole and
an extended halo with arcuate features. The subsequent spectral
monitoring at the BTA\,+\,NES has led to new results~[52]. 
In particular, a comparison of the radial velocity $V_r=-25.7$~km\,s$^{-1}$, 
obtained from numerous symmetric absorptions of a weak and moderate
intensity, with the previously published data indicates the
absence of significant velocity variations and its coincidence
with the systemic velocity based on the radio data. High spectral
resolution allowed to measure the positions of 24 rotational lines
of the (0;0) $\lambda$5165~\AA{} Swan system band of the C$_2$
molecule. As a result, an expansion velocity of the circumstellar
envelope was determined, $V_{\rm exp}=15.5$~km\,s$^{-1}$, which is
typical for stars in the post-AGB stage. In addition, the Swan
system (0;1) 5635~\AA{} band revealed a complex
emission-absorption profile. For the first time asymmetry of
strong absorption profiles of ionized metals ~(Y\,II,  Ba\,II,
La\,II, Si\,II) was detected in the optical spectrum of the source
IRAS\,23304+6347, due to the presence in these lines of a
shortwave component forming in the circumstellar envelope. The
position of the long-wave component in the profiles of these lines
corresponds to the velocity determined from the symmetric
absorptions and matches the systemic velocity. The position of the
shortwave component, on the other hand, corresponds to the
velocity based on the Swan band molecule C$_2$ and coincides with
the position of the circumstellar component of the Na\,D1 profile.
The silicon excess, the synthesis of which is possible due to the
`hot bottom process' in the hot layers of the  convective shell in
massive AGB stars, allows to talk about the belonging of the
studied object to stars with initial masses above  
4\,$\mathcal{M}_{\odot}$.

To date, the $s$-process heavy metals have been revealed in
the structured circumstellar envelopes of only three
post-AGB stars, the atmospheres of which reveal products of
the TDU. Hence, the result of the multi-year spectral monitoring
at the BTA is a discovery of a channel through which an enrichment
by the $s$-process heavy metals of structured envelopes of stars
at the post-AGB stage is taking place.

New features in the spectrum were also found as a result of
analysis of multiple observations with high spectral resolution of
a post-AGB star {\bf V448\,Lac}~[53]. A study of the behavior of 
the profiles of spectral features and the velocity field in the 
atmosphere and the circumstellar envelope allowed to detect an 
asymmetry and a temporal variability of profiles of the strongest 
absorptions with the excitation potential of the lower level  
$\chi_{\rm low}<1$~eV. Such features
are manifested primarily in the profiles of the resonant lines
Ba\,II, Y\,II, La\,II, Si\,II. The peculiarity of profiles can be
explained by a superposition of stellar absorptions and envelope
emissions. For the first time in the spectrum of V448\,Lac, an
emission was found in the Swan  system band ~(0;1)~5635~\AA\ of
the C$_2$ molecule. Based on the H$\alpha$ core, a radial velocity
variability  with an amplitude of $\Delta
V_r\approx8$~km\,s$^{-1}$ was found. The average velocity
variability  based on the weak metal lines with a smaller
amplitude $\Delta V_r\approx1$--$2$~km\,s$^{-1}$ can be a
manifestation of low-amplitude pulsations of the atmosphere. Apart
from that, time-variable differential shifts of lines in the
interval of $\Delta V_r=0$--$8$~km\,s$^{-1}$ were found. The
position of the molecular spectrum is stable in time, which
indicates the constancy of the circumstellar envelope registered
from the '$_2$ and Na\,I lines:   $V_{\rm exp}=15.2$~km\,s$^{-1}$.
The expansion velocity magnitude is typical of post-AGB stars.

For the first time the splitting of the strongest absorptions with
the low-level excitation potential of $\chi_{\rm low}<1$~eV was 
found in the optical spectra of a post-AGB star {\bf V354\,Lac}, also obtained 
with a spectral resolution of $R=60\,000$ at the 6-m BTA telescope~[54]. 
An analysis of the kinematic pattern  showed that the shortwave component 
of the split line is formed in  the powerful gas and dust envelope of
the star. Neglecting the splitting of strong lines in the
calculation of the chemical composition leads to the overestimated
excesses of $s$-process elements (Ba, La, Ce, Nd) in the stellar
atmosphere. A strong absorption profile variability was detected.
Unlike the above data for V448\,Lac, for V354\,Lac, no velocity
variation was recorded over 15 years of observations, which
suggests the stability of the velocity field in the atmosphere and
the circumstellar envelope of V354\,Lac.

\subsection{Structured envelopes of post-AGB stars}

Having obtained new results regarding the details of the chemical
composition of the sample of post-AGB stars, surrounded by complex
envelopes, we tried to find an interrelation of chemical
composition features of their atmospheres with the structural
features of the circumstellar shells. The stellar wind of the
objects at the AGB stage usually possesses a spherical symmetry,
which determines mainly the spherical shape of the slowly
expanding envelopes of these stars. However, their closest
descendants, post-AGB stars, are the objects whose central star is
surrounded, as a rule, by a non-spherical envelope of a complex
structure (a bipolar and even a quadrupole shape). A transition to
this complex morphology from the predominantly spherical shells of
AGB stars occurs exactly at the post-AGB stage, which
necessitates a detailed and comprehensive study of these objects.
A qualitatively new level of understanding the structure and
kinematics of envelopes around the far evolved stars is provided
by their observations with the Hubble Space Telescope~[51, 55]. 
These observations  have indeed revealed a bipolar structure  and jets
in many objects, previously considered to be only the point
objects. A heterogeneous asymmetric nebula RAFGL\,2688,
surrounding an F-supergiant V1610\,Cyg, faint in the
optical range, has a large spatial extent (the main features of
its structure can even be observed  with a ground-based telescope)
and shows a hierarchy of features of different scales, with
different velocities, very well illustrates the complexity of
morphology of the envelopes in post-AGB stars.

As the detailed images were accumulated, it turned out that
the so-called arcs observed in the envelopes of several
post-AGB stars indicate occasional mass-loss-rate variations
occurring at the AGB stage. The fact that the arcs are not
closed and are asymmetrical with respect to the central star,
testifies that mass loss occurred during several
episodes of wind intensification with a predominance of ejections at various
solid angles.

Based on the observations of a representative sample of stars at
the BTA with a high spectral resolution, Klochkova~[47]  considered 
the features of optical spectra of a sample of post-AGB stars with 
the atmospheres enriched in carbon and $s$-process heavy metals, and with
carbon-rich circumstellar envelopes. As a result, it was concluded
that the line profile peculiarities (the presence of an emission
component in the D-lines of a Na\,I doublet, the nature of
molecular features, asymmetry and splitting of strong absorption
profiles with a low lower-level excitation potential) are related
to kinematic and chemical properties of the circumstellar envelope
and to the type of its morphology. In particular, the variability
of the observed profiles of the absorption-emission line H$\alpha$
and lines of metals, as well as the type change
(absorption/emission) of Swan bands of the C$_2$ molecule,
observed in several objects is caused by  the variations in the
structure of the circumstellar envelope. The H$\alpha$ profile
type (pure absorption, pure emission, P\,Cyg or inverse
P\,Cyg-type, with two emission components) is not associated with
the chemical composition of the atmosphere of the central star.
The main factors, influencing the type of the H$\alpha$ profile
and its variability are the matter loss rate, stellar wind
velocity, kinematics and optical thickness of the envelope.

The above-mentioned splitting into the components of the profiles
of strongest absorptions   of heavy metals in the spectra of
supergiants V5112\,Sgr, V354\,Lac and the  central star of the IR
source   IRAS\,23304+6147 suggests that the formation process of a
structured circumstellar envelope is accompanied by its enrichment
with the stellar nucleosynthesis products. Attempts to find the
relationship between the features of the optical spectrum and the
morphology of the circumstellar environment is hampered by the
fact that the observed structure of the envelope depends heavily
on the declination of the axis of symmetry to the line of sight
and on the degree of angular resolution of the equipment, both
spectral and that used to produce direct images.

\subsection{Cool AGB supergiants}

The far evolved stars at the AGB stage, observed at the short
evolutionary phase of the transition to a planetary nebula, are
low-mass cores with a mass of about 0.6\,$\mathcal{M}_{\odot}$. Degenerate
cores are surrounded by extended gas-dust envelopes that were
formed owing to a significant loss of  matter of a star at the
previous stages of evolution. The presence of the circumstellar
gas and dust manifests itself in the peculiarities of the IR,
radio, and optical spectra of AGB. The optical spectra of these
low-mass supergiants differ from the spectra of classical massive
supergiants by the presence of molecular bands superimposed on the
spectrum of a cool supergiant, anomalies of the
absorption-emission profiles of the  H\,I, Na\,I and He\,I, lines
as well as the presence of emission lines of some metals. In
addition, all these spectral features are often variable in time.
In general, we see in the optical spectra of AGB supergiants (and
their descendants) the following features distinguishing them from
the spectra of massive supergiants: complex line profiles of
neutral hydrogen that contain the variable  in time absorption and
emission components; absorption or emission bands of molecules,
predominantly carbon bearing; envelope components of the Na\,I and
K\,I resonance lines; narrow forbidden or permitted emissions of
metals that are formed in the envelope.

{\bf R\,Sct} is a semi-regular variable of the  RV\,Tau-type.
Variable RV\,Tau-type supergiants  are in the instability band for
the population II stars. R\,Sct is the brightest and unstable
representative of these stars, has a  variability period of about
142~days. Kipper and Klochkova~[56], using the R\,Sct spectra, obtained 
at the BTA\,+\,NES and also using the spectra from the archive of the 
NARVAL spectropolarimeter of the Peak~du~Midi observatory TBL telescope, 
determined the fundamental parameters, studied the characteristics of the 
spectrum and the kinematic pattern, and also calculated a detailed chemical
composition of the atmosphere. The metallicity of the star is
${\rm Fe/H}\approx -0.5$, at a very large excess of carbon  ${\rm
[C/Fe]}=+0.84$. No excess of the $s$-process heavy metals was
detected, therefore an assumption was made that the star has not
yet experienced the TDU. The resulting effective temperature
T$_{\rm eff}=4500$~K also indicates that R\,Sct is still at the AGB stage.

The spectrum of R\,Sct revealed the splitting of the strongest absorptions
as well as the presence of weak and variable emissions in the  Fe\,I,
Ti\,I lines~[56]. Near the brightness minimum, emission is also observed in
the H$\alpha$, H$\beta$ lines.
Radial velocity measured from the weak symmetrical absorption, varies with a large amplitude.
The pattern of radial velocities  measured from
different types of spectral features varies with time, and a velocity stratification
is observed for selected phases.

A cool {\bf supergiant V1027\,Cyg} is associated with an IR source
IRAS\,20004+2955. Kwok~[57]  included this star in his now widely known list 
of candidates in protoplanetary nebulae. However,  V1027\,Cyg so far remains 
poorly studied, there is no consensus on the evolutionary status of this star. Arkhipova
et al.~[58]  based on the UBV-photometry and low-resolution spectroscopy classified 
this star as a semi-regular variable with the brightness variation amplitude  increasing from
the $V$ band to $U$ band from $0\fm4$ to $0\fm8$. Later, measuring
radial velocities with a correlation spectrometer, Arkhipova et
al.~[59]  found a radial velocity variability in the range from 
$-10$ to $+20$~km\,s$^{-1}$ and noted a strengthening  of the Ba\,II lines. Klochkova et
al.~[60], using the spectra taken with the echelle spectrographs of the 6-m BTA telescope, 
and applying the method of model atmospheres determined the main parameters of
V1027\,Cyg (effective temperature  $T_{\rm eff}=5000$~K, surface
gravity $\log g=1.0$), metallicity ${\rm [Fe/H]}_{\odot}=-0.2$ and the 
abundances of sixteen elements in the atmosphere of the star. A close to solar 
metallicity combined with a modest radial velocity of the star suggest that 
V1027\,Cyg belongs to the Galactic disk population.

A more detailed study of the characteristics of the spectrum and
the velocity field in the atmosphere of V1027\,Cyg was carried out
based on the high spectral resolution observations   with the NES
echelle spectrograph of the 6-m telescope~[61]. From the symmetric
absorptions of a low and moderate intensity, slight variations in
radial velocity $V_r$ (abs) were revealed with an amplitude of
about 5~km\,s$^{-1}$ occurring due to pulsations. In the red
spectral region,  numerous weak lines of the CN molecule were
identified and the K\,I\,7696~\AA{} line, which has a P\,Cyg-type
profile. An agreement of   radial velocities, measured from the
symmetric absorptions of metals and from the CN lines, indicates
the formation of the CN spectrum   in the stellar atmosphere.
Numerous interstellar bands, DIBs, were identified in the
spectrum, the position of which in the spectrum,  
V$_r{\rm(DIBs)}=-12.0$~km\,s$^{-1}$,  corresponds to the  velocity of the
interstellar medium in the Local Arm of the Galaxy.

The spectrum of V1027\,Cyg reveals a long-wave and temporally
variable H$\alpha$ profile shift due to the distortion of the core
and the shortwave line wing. The splitting of the strongest metal
absorptions and their ions (Si\,II, Ni\,I, Ti\,I, Ti\,II, Sc\,II,
Cr\,I,  Fe\,I, Fe\,II, Ba\,II) was discovered in the spectrum of
the star for the first time. A broad profile of these lines
contains in its core a stably located weak emission, the position
of which can be considered as the systemic velocity $V_{\rm
sys}=5.5$~km\,s$^{-1}$. A more specific interpretation of the
radial velocity pattern,  isolation of the  circumstellar and
photospheric components on the   Na\,I D-line  profiles, and,
consequently, a refinement of the distance to the star, and
defining its evolutionary status yet requires spectroscopy with
{\it extremely high} spectral resolution.

{\bf IR source RAFGL\,5081} (IRAS\,02441+6922) was registered in
the earliest balloon IR surveys. Later, Kwok et al.~[62]  added 
this source to the list of candidate objects in the  AGB stars.
Currently, there is no information about the central star of the
RAFGL\,5081 source. The SIMBAD database  contains no observational
data for this object, both spectral and photometric, in the
visible wavelength range. Publications about the observations of
the IR source RAFGL\,5081 in other wavelength ranges are rather
scarce. After completing the infrared observations of RAFGL\,5081
at the CFHT, having adopted the data of the IRAS satellite, Kwok et al.~[62]  
attributed this object to a group of sources with a maximum IR radiation 
in the region  of $\lambda < 10~\mu$, but could not say anything sure about 
its status. No radiation in the CO, OH, H$_2$O bands  is registered~[63, 64].

\begin{figure}[ht!]
\includegraphics[scale=0.6,  bb=40 65 475 530,clip]{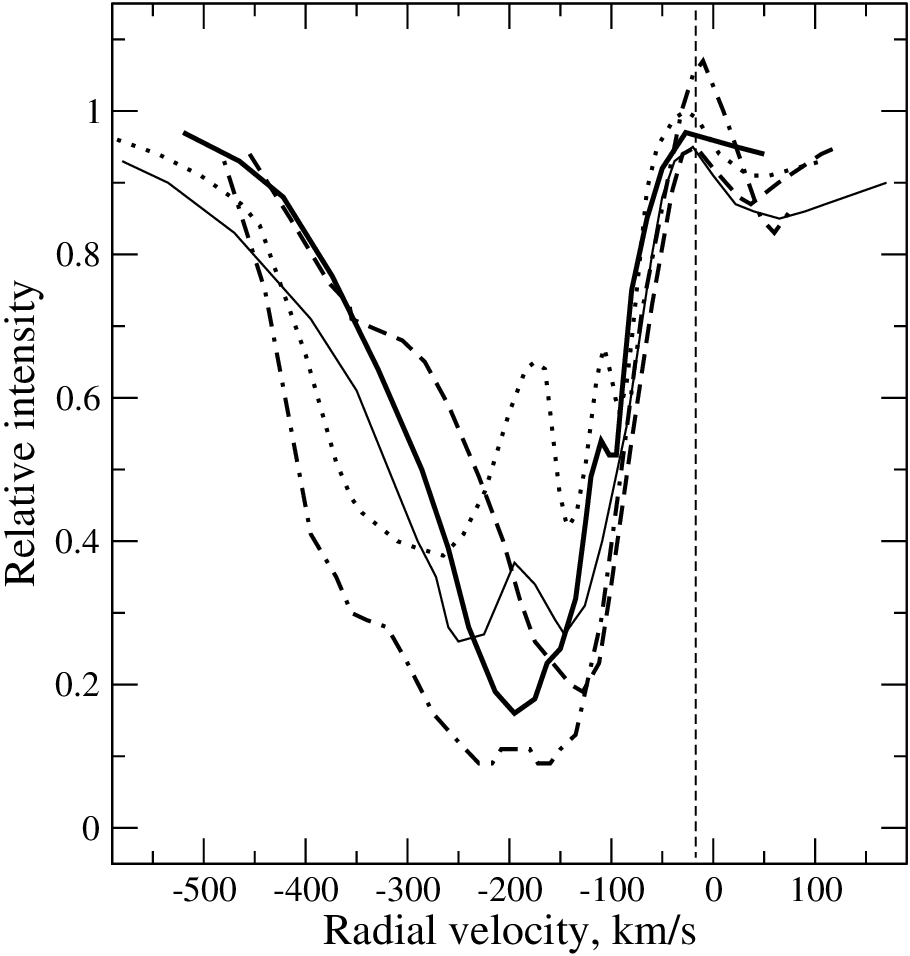}
\caption{An example of variability of the H$\alpha$ profile in
the selected spectra of RAFGL\,5081: the thin solid line -- March 7, 1999,
the dot-and-dash line -- December 1, 2001, the dashed line -- February 22,
2003, the dotted line -- January 14, 2006, the continuous solid line -- February 3,
2007. The vertical dashed line describes the systemic
velocity V$_{\rm sys}=-21$~km\,s$^{-1}$ from~[65].}
\label{RAFGL5081_Halpha}
\end{figure}

The optical spectra of the weak in the visible  region central
star of the IR source   RAFGL\,5081 were for the first
time studied~[65]  based on a long-term spectral monitoring with 
high spectral resolution conducted at the 6-m telescope with the NES 
spectrograph over 2001--2015. From a comparison with related objects 
the spectral class of the star is determined to be close to G5-8\,II, 
what is consistent with the effective estimate of the temperature of the star 
T$_{\rm eff}=5400\pm100$~K, which is made according to spectral criteria.
From the equivalent width of the oxygen triplet  O\,I\,7773~\AA{}
W(O\,I)\,=\,0.84~\AA{}  in the spectrum of the companion of
RAFGL\,5081, a preliminary luminosity estimate  is done:  
M$_{\rm bol}=-4^{\rm m}$. A low luminosity allows to exclude the option of
a massive red supergiant. In the spectrum of the companion of
RAFGL\,5081 an absorption Li\,I~6707~\AA{} with the equivalent
width of $W\approx 0.1$~\AA{} is reliably identified, which
indicates that the object belongs to more massive AGB stars with
an initial mass of $M_{\rm init} \ge 4~\mathcal{M}_{\odot}$.

An unusual spectral phenomenon has been detected in the spectra of
the star: broad absorption profiles of medium and low intensity
are splitted and, unlike the wind absorptions, are stationary. The
stationarity of absorptions eliminates the possibility of
explaining the anomalous  profiles by the spectroscopic binarity
of the star. Radial velocities for the wind components of the
Na\,I  D-line and H$\alpha$  profiles reach the  values of
$-250$~km\,s$^{-1}$ and $-600$~km\,s$^{-1}$, respectively
(see Fig.~4). These profiles contain narrow components whose
number, depths and positions vary with time. The temporally
variable multicomponent structure of the  Na\,I D-line, and
H$\alpha$ profiles indicates non-uniformity and instability of the
circumstellar envelope of RAFGL\,5081.

For all dates of observations, heliocentric radial velocities
$V{\rm r}$ were measured corresponding to the positions of all
components of metal absorptions, as well as those of the Na\,I
D-lines and H$\alpha$. An analysis of multicomponent profiles of
Na\,I\,(1) lines revealed the presence of components with the
velocity of V$_r\rm {IS}=-65$~km\,s$^{-1}$, based of
which it was concluded that RAFGL\,5081 is located behind the
Perseus Arm, i.e. not closer than 2~kpc.

\subsection{Detection of spectral binarity (SB2) of the star BD$-$6\,$\degr$1178}

A poorly studied cool star  BD$-6\degr1178$ is identified with
IRAS\,05238$-$0626, an IR-source at the galactic latitude of
$b\approx-22\degr$. Due to the presence of a  radiation excess in
the region of 12--60~$\mu$ and taking into account the position on
the IR color diagram, this object is considered to be a candidate
in protoplanetary nebulae~[66]. Some signs of photometric variability 
were found  for the object~[67]: the average magnitude in the
system close to the $R$ filter varies in the $[10\fm78;10\fm87]$
interval with an average error of about $0\fm01$. As for the
spectroscopy of BD$-6\degr1178$, before the observations at the
BTA, the literature contained only the data about the
low-resolution spectra, approximately 5~\AA/pixel, based on which
the spectral class was estimated as
F2\,II~[68]. From the  high spectral resolution spectra obtained with 
the BTA\,+\,NES on arbitrary dates, spectral classification was carried out, 
and   absorption profiles  and the velocity field in the atmosphere of the  star
were thoroughly studied~[69]. The results of the quantitative spectral 
classification of the star BD$-6\degr1178$ led to a conclusion on the 
spectroscopic binarity of the star. Both components are F stars, F5\,IV--III\,+\,F3V,
with close rotation velocities: 24 and 19~km\,s$^{-1}$. For four times of 
observations  heliocentric radial velocities of both components of the binary 
system were measured. This is a double-lined spectroscopic binary with  
rather narrow and well resolved lines. The maximum recorded mutual shift of the
component spectra is about 120~km\,s$^{-1}$.

In general, no grounds were found for the classification of BD$-6\degr1178$
as a post-AGB star. The coordinates of BD$-6\degr1178$ and the distance to
it of $d\approx450\pm37$~pc suggest its belonging to the Ori\,OB1 association.
Therefore, BD$-6\degr1178$ may be a young object of the Galactic disk on
the stage prior to MS. Note that earlier Torres et al.~[70], exploring a sample 
of T\,Tau-type candidate stars, included in it BD$-6\degr1178$ as well.
Based on the analysis of spectral and photometric data these authors found 17 
new T\,Tau-type stars  and 13 new Ae/Be Herbig-type stars.
However, they were unable to include BD$-6\degr1178$ in any of these groups,
attributing it to the mixed-type object group (``miscellaneous'').

Detecting spectroscopic binarity of the SB2 type also gives us
reason to doubt the classification of BD$-6\degr1178$ as a
post-AGB star, since among the known stars on this evolutionary
stage there are no SB2 binaries. A number of post-AGB stars are
binary systems, but of SB1-type. The nature of their invisible
companion is unknown, since its features are not observed in the
spectra of binary post-AGB stars.  Either a white dwarf, or
a low-mass MS star can be a companion.

The features of post-AGB stars in the binary systems are presented
in van~Winckel's surveys~[71, 72]. Note that as a result of a long-term 
monitoring, Hrivnak et al.~[73]  did not find any binary systems among the 
PPN with a 21~$\mu$ feature and with the atmospheres enriched with heavy metals.

\subsection{Stars approaching the PN stage}

Our program also includes a number of hot stars, which have closely 
approached the stage of a planetary nebula. An IR source {\bf IRAS\,01005+7910} 
(hereinafter referred to as IRAS\,01005) located high above the Galactic plane, 
its latitude is ${b = 16\fdg6}$. In the optical range, the object is associated
with a nameless peculiar B-supergiant.  The source's position on
the IRAS color diagram is consistent with the stage of the
protoplanetary nebula. Photometric variability of IRAS\,01005 was
studied by several groups. Hrivnak et al.~[74]  have noted the brightness
variability  of an object on a  scale which is very short for the
protoplanetary nebulae, less than a few days. Arkhipova et al.~[75]  
during the long-term photometric monitoring conducted in the UBV-bands 
for several hot objects, including IRAS\,01005, found a fast and irregular 
variability of their brightness. Based on these observations of a low-amplitude
variability, the authors of~[75]  came to a conclusion on the variations in 
the stellar wind parameters of these stars and (or) about the presence of 
micro pulsations with characteristic periods of several hours.

\begin{figure}[ht!]
\includegraphics[scale=0.6]{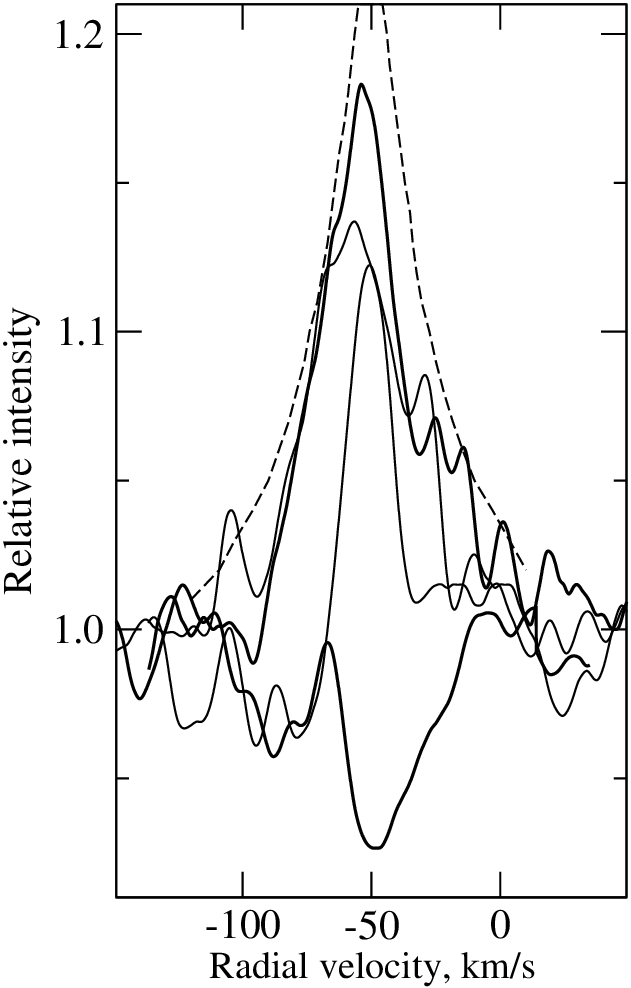}
\caption{The Si\,II profile variations  in the spectrum of IRAS\,01005:
{\it 1} -- the Si\,II\,(2) 6347 emission (May 29, 2013); {\it 2} and {\it 3} -- 
 the Si\,II\,(2) 6347~\AA{} profiles, obtained on April 13,
2003 and November 15, 2005, respectively; {\it 4} -- the average of
the emission-absorption profiles of Si\,II\,(3) 4128 and 4130~\AA{}
(May 29, 2013). The dashed line --the external envelope of
all Si\,II 5979~\AA{} profiles. The intermittent vertical line -- the
accepted systemic velocity. A color image is given in the
electronic version of the article. V$_{\rm sys}\approx 50.5$~km\,s$^{-1}$.  }
\label{IRAS01005_Si}
\end{figure}

The principal results for the central star of  IRAS\,01005 are obtained in 
the paper~[76], whose authors, by defining the fundamental parameters of this 
supergiant (T$_{\rm eff}=21\,500$~K,  surface gravity $\log g =3.0$, metallicity 
${\rm [Fe/H]}=-0.31$ and  abundance of a number of chemical elements), confirmed for it 
the   post-AGB stage. 
A high galactic latitude combined with reduced metallicity indicates that
the IRAS\,01005 belongs to the old population of the Galaxy. An
important result of this work is a detection of a carbon excess
($\rm C/O>1$) in the atmosphere of the central star in the IR spectrum 
of the circumstellar envelope of which spectral features of a carbon-containing 
molecule, fullerene C$_{60}$ were detected~[77].

After detecting the variability of the spectrum  of IRAS\,01005 at
the 6-m telescope,  its central star was monitored with high
spectral resolution,  $R=60\,000$. Based on 23 spectra, the
spectral class of the central star was determined as ${\rm B}1.5\pm0.3$,  
the luminosity class Ib,  numerous spectral features were identified, including 
many forbidden ones, the variability of their profiles and radial velocity was
considered~[78]. Based on the position of symmetric and stable forbidden 
emission profiles [N\,I], [N\,II], [O\,I], [S\,II] and [Fe\,II],  systemic velocity 
V$_{\rm sys}=-50.5$~km\,s$^{-1}$ was determined. The presence of forbidden
emissions [N\,II] and [S\,II] indicates the onset of ionization of
the circumstellar envelope and the proximity of the planetary
nebula stage. The differences in radial velocity V$_r$ determined
from the line cores, of about 34~km\,s$^{-1}$, are partly caused
by the deformations of profiles by variable emissions. The V$_r$ difference 
from the wings of lines is smaller, it is approximately equal to 23~km\,s$^{-1}$, 
and may be due to the pulsations and/or a hidden star binarity. 
Profile deformations of the absorption-emission lines may be associated 
with the variations in their absorption (photospheric) components with  variations of
geometry and kinematics at the base of the wind (see Fig.~\ref{IRAS01005_Si}). 
Our material suggests that over two days the velocity variations  reach a confidently
measured value.

\begin{figure}[ht!]
\includegraphics[angle=0,width=0.4\textwidth]{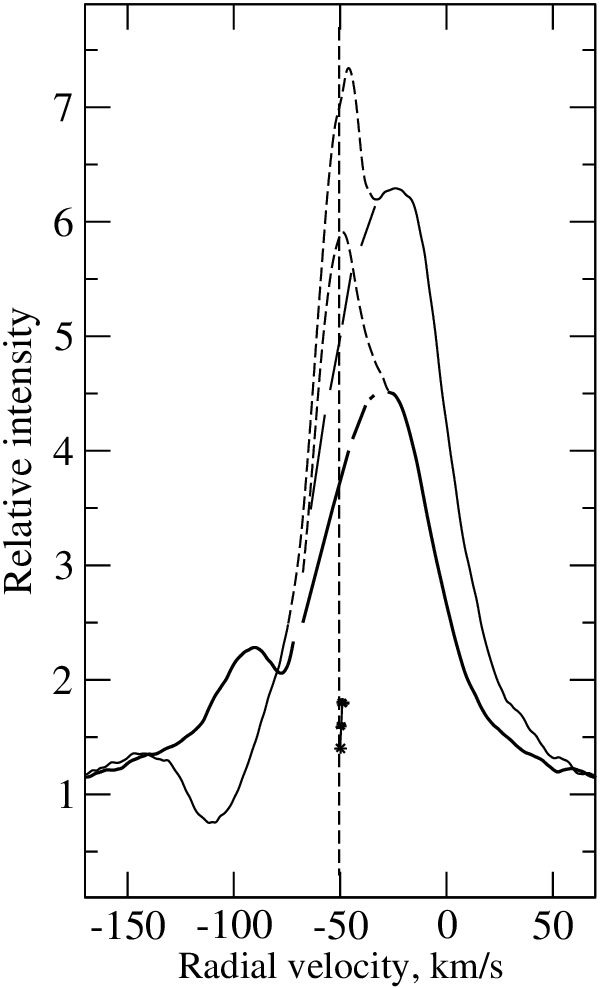}
\caption{The H$\alpha$ profiles in the spectrum of IRAS\,01005,
obtained on January 18, 2005 --{\it 1} and May 9, 2004 -- {\it 2}.
The strokes -- envelope emissions, the dash lines -- wind profile
sections underneath them. A vertical intermittent straight line
describes the radial velocity for forbidden emissions, a chain of
crosses next to it is the bisector for the lower part of the
profile of January 18, 2005.} 
\label{IRAS01005_Halpha}
\end{figure}

As can be clearly seen in Fig.~\ref{IRAS01005_Halpha}, H$\alpha$ lines
in the spectrum of the central star IRAS\,01005 possess P\,Cyg\,III0-type wind profiles.
It is shown that the deviations of the wind from the spherical
symmetry are small. We recorded a variable wind velocity (in
the interval of 27--74~km\,s$^{-1}$ for different times of observations) and a
high intensity of the long-wave emission (the continuum level in excess up to seven times),
which is not typical for the classical supergiants, but rather for hypergiants.
For the five main components of Na\,I D-lines the heliocentric radial velocities are:
$V_r=-72.5$, $-65.3$, $-52.2$, $-27.7$ and $-10.2$~km\,s$^{-1}$,
which, within the error limits, coincides with the  published data.  A weak component,
$V_r=-52.2$~km\,s$^{-1}$, is formed in the stellar atmosphere, two
more long-wave  components are interstellar, forming in the Local Arm.
The presence of the $V_r=-65.3$~km\,s$^{-1}$ component,
apparently originating in the  interstellar medium of the
Perseus Arm, allows us to consider $d=2.5$~kpc as the
bottom distance estimate to IRAS\,01005.
The most shortwave component, $V_r=-72.5$~km\,s$^{-1}$, can be formed in the
circumstellar envelope expanding with the velocity $V_{\rm exp}\approx 22$~km\,s$^{-1}$ 
typical for the PPN.

\begin{figure*}[ht!]
\includegraphics[scale=0.55,  bb=40 40 750 530,clip]{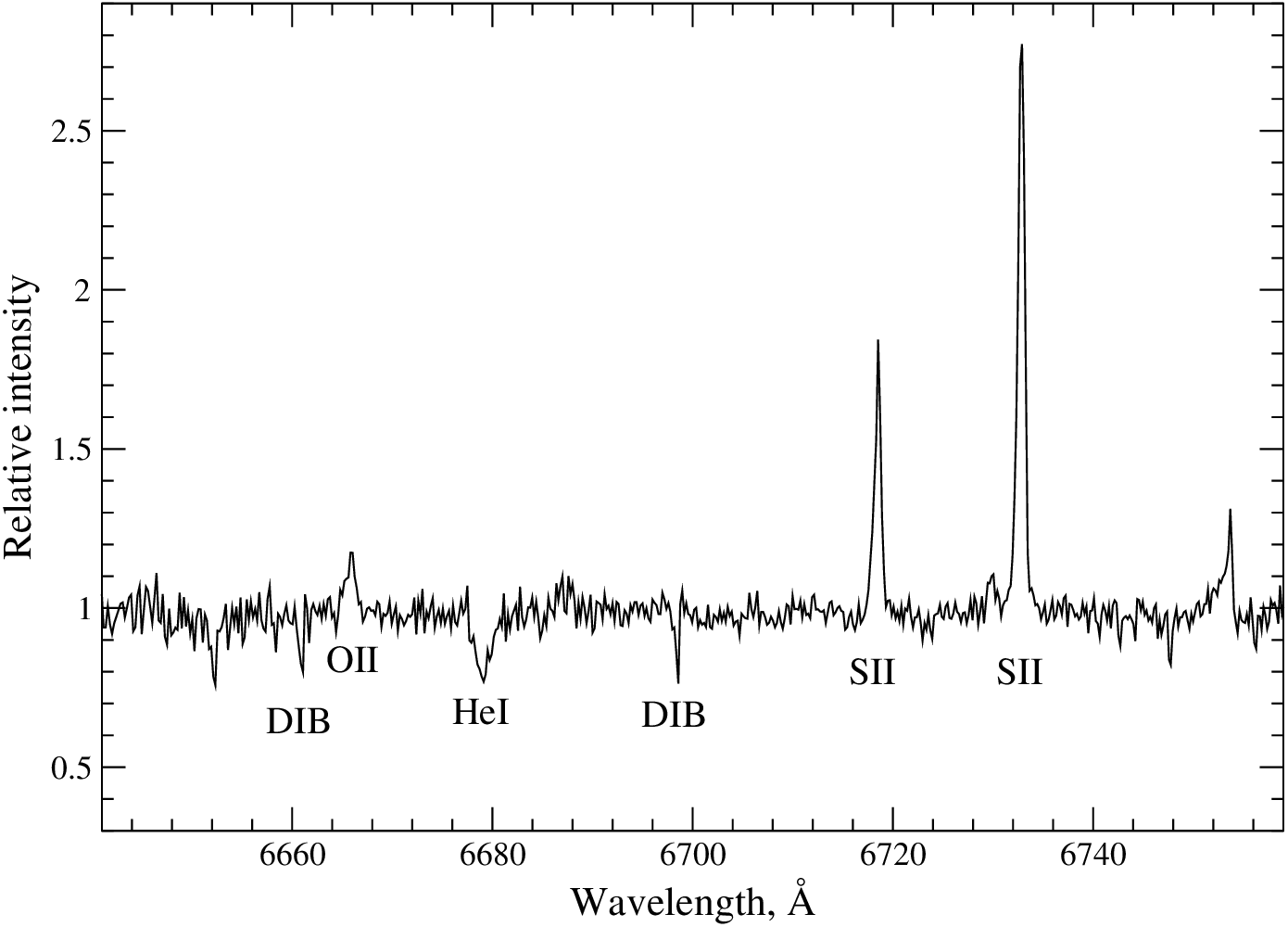}
\caption{A fragment of the spectrum of StH$\alpha$ 62 in the interval containing the [S\,II]  
  $\lambda$ 6717 and  6730~\AA{} forbidden lines. The main features of the fragment are identified.} 
\label{IRAS07171_SII}
\end{figure*}

In cooperation with a group of specialists from the Moscow University Sternberg 
Astronomical Institute a complex research (long-term photometry, spectroscopy of
a moderate and high resolution) was conducted of several related hot
stars, close to the PN phase:  V1853\,Cyg (IRAS\,20462+3416)~[79],  
V886\,Her (IRAS\,18062+2410)~[80],  StH$\alpha$62 (IRAS\,07171+1823)~[81].
Photometric parameters of these objects vary with an amplitude
approximately equal to $0\fm2$--$0\fm3$. Their optical spectra are
somewhat different, due to a slight difference in the degree of
approximation to the planetary nebula phase. The spectrum of the
B-supergiant V886\,Her includes complex H\,II  lines with powerful
emission components, light metal ion absorption lines, forbidden
emissions, as well as a rich nebular spectrum of the envelope~[80].

The same methods were used to study the properties of a high-latitude planetary nebula  
G\,199.4+14.3~[81].  The spectrum of  its central star StH$\alpha$62 in the range of $\lambda$~4330--7340~\AA\ 
contains absorption lines, corresponding to the spectral class Sp\,=\,B0.5, 
in combination with the emission lines of the gas envelope.
Figure~\ref{IRAS07171_SII} in a way of  example demonstrates  a fragment of 
the StH$\alpha$62 spectrum obtained at the 6-m telescope with the PFES spectrograph~[48], 
containing the [S\,II] $\lambda$~6716.45 and 6730.85~\AA{} forbidden lines. The presence
of the [N\,II] and [S\,II] forbidden emissions  allowed the
authors of~[81]  to estimate the density and temperature of the gas envelope  
of the star, the values of which are usual for the low excitation planetary 
nebulae within the range of formation of the [S\,II] and [N\,II] lines, where the
gas ionization is just starting. The rate of expansion of the
outer shell from the forbidden lines amounted to 12--13~km\,s$^{-1}$. 
Assuming that the star is a low-mass  ($\mathcal{M}=0.55\pm0.05~\mathcal{M}_\odot$) 
protoplanetary object, a  distance estimate of $d=5.2\pm 1.2$~kpc was found. As shown
in~[81], the main characteristics and features of the optical spectrum of 
V1853\,Cyg are close to those of StH$\alpha$62.

{\bf HD\,331319} is a  post-AGB candidate, an optical component of
the IR source IRAS\,19475+3119. New results for it were for the
first time obtained with the echelle spectrograph of the 6-m
telescope. The fundamental parameters and a detailed chemical
composition of its atmosphere were determined using the model
atmosphere method~[82]. A particularly significant  result is  
detection of helium lines in the spectrum of this object with an 
effective temperature   T$_{\rm eff}=7200$~K. An excess of helium 
in the observed layers of the atmosphere is interpreted as a consequence 
of the synthesis of this element during the previous evolution of the star. 
An excess of nitrogen and oxygen is also detected: 
${\rm [N/Fe]}_{\odot}=+1.30$~dex, ${\rm [O/Fe]}_{\odot}=+0.64$~dex at a
slight excess of carbon. The metallicity of the stellar atmosphere
${\rm [Fe/H]}_{\odot}=-0.25$   is slightly different from the solar value. 
The  $s$-process metals abundance is underestimated relative to the metallicity:  
for Y, Zr  ${\rm [X/Fe]}_{\odot}=-0.68$. The barium abundance is also
underestimated with respect to metallicity: ${\rm
[Ba/Fe]}_{\odot}=-0.47$. The Li 6707~\AA{} line was not found in
any of the obtained spectra of the star. In general, the abundance
of chemical elements raises doubts about the belonging of the
object IRAS\,19475+3119 to the post-AGB evolutionary stage.
Metallicity combined with the radial velocity V$_r=-3.4$~km\,s$^{-1}$ and 
galactic latitude $|b|=2\fdg7$ of the object indicate its belonging to the 
population of the disk of the Galaxy. According to the position of the 
absorption bands formed in the circumstellar envelope, the expansion 
velocity of the shell was determined, V$_{\exp}\approx 21$~km\,s$^{-1}$.

\subsection{Spectral atlases}

Working with high quality spectral material we obtained at the BTA
for the supergiants of various types (different spectral classes,
luminosity classes,   evolutional stages, chemical composition
features), we considered creation of spectral atlases as a
necessary stage of our work. The following atlases, relevant for
this context are published: the spectra of hypergiants and supergiants~[83], 
the spectra of a canonical post-AGB star HD\,56126~[84].
The atlas~[85] presents a comparison of spectra of a low-mass post-AGB 
supergiant BD$+48\degr1220$ (IRAS\,05040+4820), a peculiar supergiant 3\,Pup
(IRAS\,07418$-$2850) and a massive supergiant $\alpha$\,Cyg (=IRC+50337).

A detailed study of the spectrum of a post-AGB supergiant
BD$+48\degr1220$ was done in~[86], the authors of which, applying the method 
of model atmospheres determined the main parameters of the star and the chemical 
composition of its atmosphere, slightly different from that of the  Sun.
Given a small excess of the iron peak elements, ${\rm [Met/H]}_{\odot}=+0.20$,
a mild excesses of carbon and nitrogen was found ${\rm [C/Fe]}_{\odot}=+0.25$, 
${\rm [N/Fe]}_{\odot}=+0.27$. The $\alpha$-process elements Mg, Si and Ca have 
a small excess, on the average of $[\alpha/{\rm H}]_{\odot}=+0.12$. A large excess 
of sodium was revealed, ${\rm [Na/Fe]}_{\odot}=+0.75$, which is probably a consequence of
the outflow of matter, processed in the NeNa cycle, into the atmosphere.

Obviously, the spectral atlases published in the accessible
form, contribute to the preservation of observational data and undoubtedly
facilitate the subsequent work of spectroscopists, including the beginners.
Our atlases contain spectral material both in the graphical representation,
as well as tables with a detailed identification of all spectral features. 
A major part of these atlases is available in digital form on the Internet.

\subsection{Interstellar and circumstellar features}

A fundamentally new result was also obtained in relation to the
formation of mysterious spectral features identified with the
so-called diffuse interstellar bands~(DIBs). It turned out that in
the spectrum of V5112\,Sgr the radial velocity determined from the
DIBs coincides with the velocity from the shortwave envelope
component of the Na\,I D-lines, which allowed to conclude
on the formation of  DIBs in the circumstellar
envelope~[50]. There are several programs aimed at finding the DIBs in 
the circumstellar envelopes, but all attempts were unsuccessful: the discovered 
envelope features were not subsequently confirmed. Based on the extensive and
high-quality spectral material providing the high accuracy of
positional measurements in the spectra, as well as having an
extensive experience with the spectra, we have for the first time
obtained a reliable positive result in the long-term search for
the DIB analogues in the circumstellar medium.

Note that Klochkova et al.~[52]  measured the positions of five 
absorptions, which can be identified with the DIBs in the spectrum of 
the central star of IRAS\,23304+6147, namely:  5797, 6196, 6203, 6207, 6613~\AA. 
The average velocity value for them is V$\rm _r(DIBs)=-15$~km\,s$^{-1}$. 
If we discard the 6613~\AA\ band, which is blended in the spectrum of
the studied supergiant with a strong  Y\,II line,  we then obtain
an average velocity of V$\rm_r(DIBs)=-14.0\pm 1.3$~km\,s$^{-1}$,
close to the velocities measured from the longest-wave component of the Na\,I D-lines, 
V$_r=-13.2$~km\,s$^{-1}$. 
Thus, we conclude that in the case of IRAS\,23304+6147 the DIBs are formed in 
the interstellar medium. Hence, the star V5112\,Sgr still remains the only star at the
post-AGB stage, in the spectrum of which the DIBs, forming in the
circumstellar envelope  are found~[50].

\section{SPECIAL REPRESENTATIVES OF POST-AGB-STAR FAMILIES}

Let us briefly discuss some of the special types of stars observed
at the post-AGB stage. The first step for us in learning the
post-AGB candidates were the F-supergiants at high galactic latitudes. 
The  prototype star for them is a peculiar supergiant UU\,Her. In the 1980s 
these strange objects received a lot of attention, which was due to the 
internal contradiction of the original idea about them as the classical supergiants residing
at such high latitudes of the Galaxy. However, the subsequent
study of the chemical composition and other observable properties
of these stars showed that both UU\,Her, and related objects are
in fact the far evolved {\bf low mass} stars belonging to the old
population of the Galaxy. The contradiction between the high
luminosity of UU\,Her-type stars and their location at high
latitudes of the Galaxy has thus been resolved.

\begin{figure*}[ht!]
\includegraphics[scale=0.5,  bb=40 40 710 530,clip]{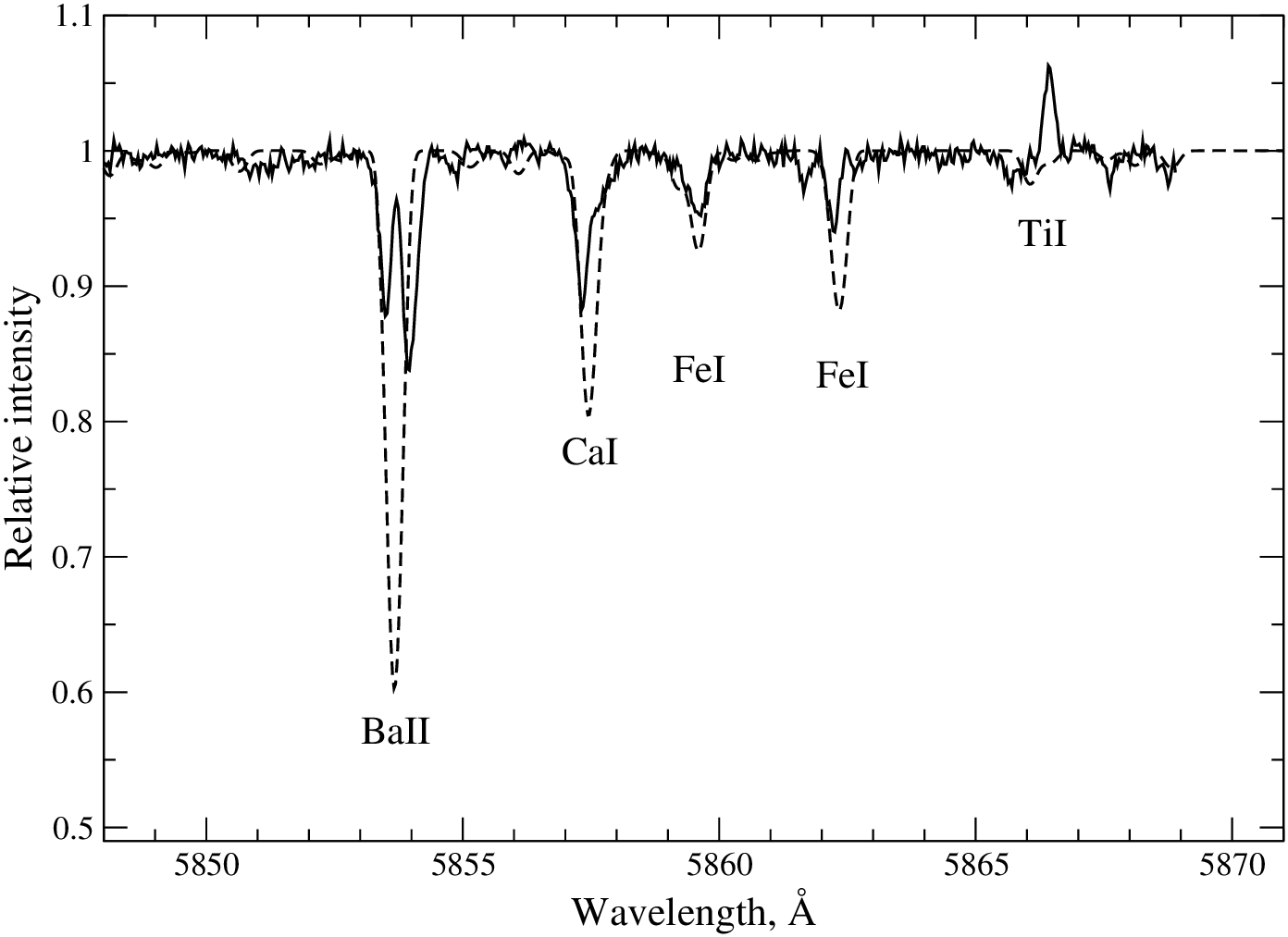}
\caption{A fragment with the Ti\,I $\lambda$~5866.40~\AA{}
emission in the
spectra of LN\,Hya, obtained in a quiet (April 2, 2010, described by the
dashed line) and in the excited (June 1, 2010---the solid
line) states of the atmosphere of the star. The identification of the
main spectral lines of this fragment is indicated.} 
\label{LNHya_sp}
\end{figure*}

A high-latitude {\bf supergiant  LN\,Hya} belongs to the
semi-regular variables that are the far evolved stars in the
post-AGB stage. The observations of LN\,Hya, carried out at the
6-m telescope with a high spectral resolution in 2003--2011,
allowed us to study in detail the features of its optical spectrum
and the velocity field in the atmosphere~[87].
Radial-velocity variations from date to date with an amplitude of
up to 3\,km\,s$^{-1}$ were found from the weak symmetric
photospheric absorptions,  which is a consequence of weak
pulsations. For a long time, an emission in H$\alpha$ was
considered to be the only peculiar feature of the LN\,Hya
spectrum. However, as a result of spectral monitoring with a high
spectral resolution we discovered a peculiarity and variability of
the profiles of strong lines of Fe\,I, Fe\,II, Ba\,II, Si\,II, and
others. For the first time an asymmetric shape of the profiles of
these lines was revealed: their shortwave wings are elongated, and
the nuclei are either split or distorted by emission. The
observational season of 2010 was the most intriguing, when from
one spectrum to another the position and depth of the H$\alpha$
absorption component, the intensity of the shortwave and longwave
emission components, as well as the ratio of their intensities had
been varying. In the spectrum of June 1, 2010, there appeared weak
emissions of neutral atoms  (V\,I, Mn\,I, Co\,I, Ni\,I, Fe\,I),
which is clearly seen in the spectral fragment, presented in
Fig.~\ref{LNHya_sp}. These features of the stellar
spectrum, registered for the first time give us grounds to assume
that in 2010 we managed to register short-term variations in the
physical state in the upper layers of the LN\,Hya atmosphere.

\subsection{Close binaries with a large  hydrogen deficiency}

A special subgroup of stars at the post-AGB stage is a very small
group of binary systems with a large hydrogen deficiency (HdBs).
The prototype of these close binaries is a bright enough star
$\upsilon$\,Sgr (HD\,181615). Only  four HdBs-type stars are known
up to date, and these objects possess very close effective
temperatures, about 10\,000~K. It is assumed that the post-AGB
companion in these binary systems is in the helium burning stage
in the layer surrounding the degenerate C-O--core. The
lifetime of the helium supergiant is only about $10^4$~years,
which explains the rare occurrence of HdBs stars. The nature of
the second, invisible system component remains unknown. A keen
interest to these objects is due to the fact that, according to
modern concepts, they are considered to be the predecessors of the
SN\,Ia supernovae.

\begin{figure*}[ht!]
\includegraphics[scale=0.5, bb=40 50 700 530,clip]{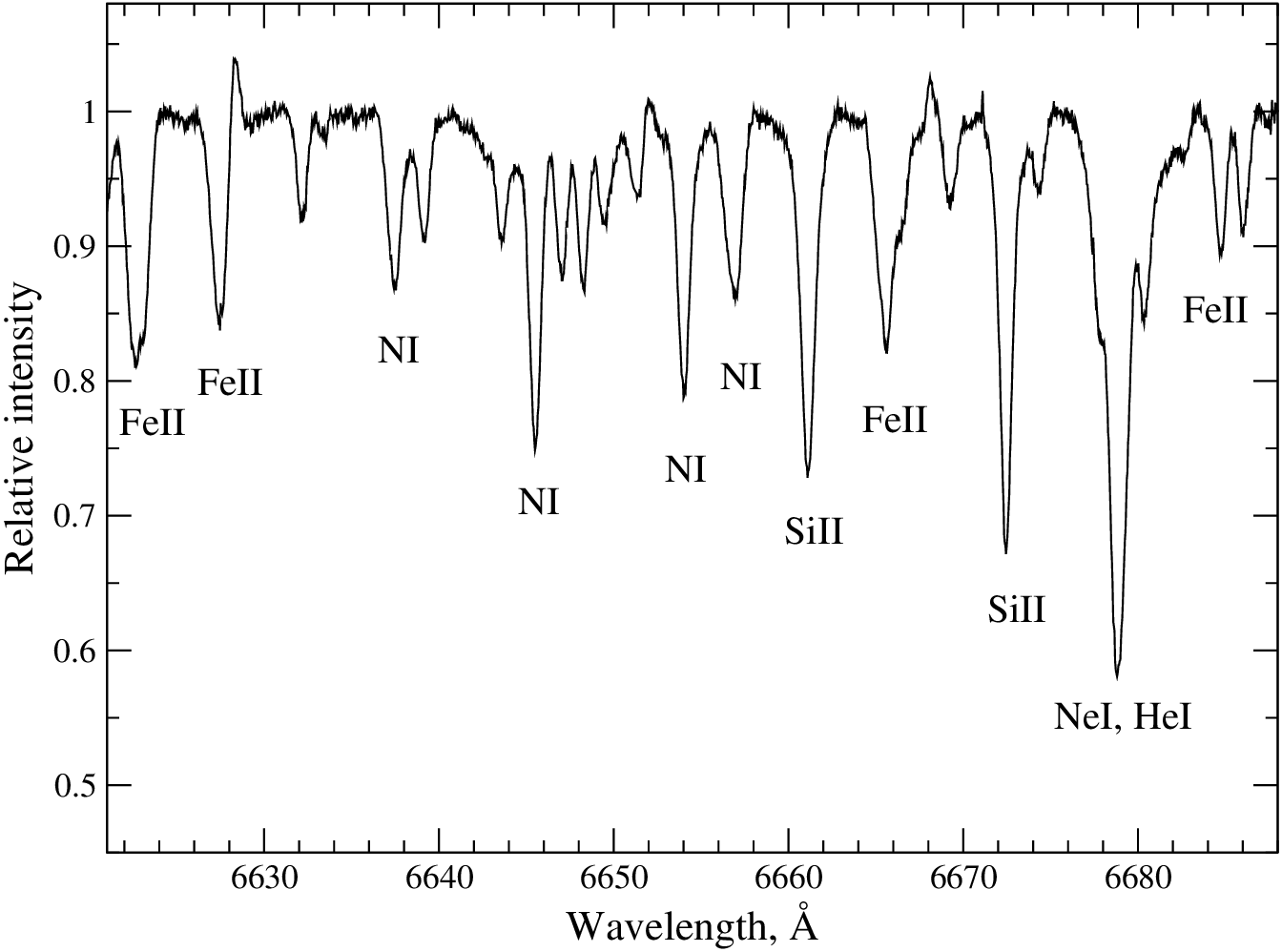}
\caption{A fragment of the spectrum of $\upsilon$\,Sgr with strong nitrogen  lines. 
       The main spectral lines of this fragment are identified~[88].}
\label{Ups_Sgr}
\end{figure*}

Currently, the {\bf  $\upsilon$\,Sgr and KS\,Per} A-supergiants are very well studied.
The main features of the supergiant $\upsilon$\,Sgr (Sp\,=\,A2\,Ia) are a strong and variable
emission in H$\alpha$ and a large excess of the IR flux.
According to the spectra obtained at the BTA with the NES spectrograph, using the data from
the archive of the Pic du~Midi Observatory 2-m telescope, Kipper and
Klochkova~[88]  determined a detailed chemical composition
of the supergiant's atmosphere in the $\upsilon$\,Sgr system.
The hydrogen abundance is reduced by five orders of magnitude
with a slight iron deficiency ${\rm [Fe/H]}=-0.8$~dex.
Interestingly,  a similar hydrogen deficiency was obtained for the star KS\,Per
(IR source IRAS\,04453+4311).
The atmospheres of both stars revealed a large nitrogen excess and
a noticeable excess of heavy metals. Figure~\ref{Ups_Sgr}
presents a fragment of the spectrum of $\upsilon$\,Sgr, saturated with the anomalously
strong lines of the  N\,I atom. A rare for this stage of evolution
feature of the $\upsilon$\,Sgr spectrum  are numerous
emissions of metals and their ions, the place of formation of which can
be an envelope around the binary system or the disk.

\subsection{Stars at the late helium flash stage}

A famous supergiant FG\,Sge over the past 100 years shows a high
rate of evolution: throughout the live of one generation of
observers, the star has crossed over the Hertzsprung--Russel
diagram. Bloecker and Schoenberner, modeling the evolutionary
behavior of stars of small and intermediate masses, devoted a
separate study~[89]   to the evolution of FG\,Sge. We refer 
the interested readers  to this publication itself, without 
considering it in detail here. Van~Genderen and Gautschy~[90], 
having gathered the FG\,Sge photometry data  over 100 years of 
observations, restored the course of its evolution from the state 
of a hot star Sp\,=\,O3 in 1880 (the central star of a planetary 
nebula  He\,1--5) up to a cool star of spectral class K2 in 1992. 
Given that, the radius of the star increased   more than two orders
of magnitude from 1 to 184~$R_{\odot}$. Iben and MacDonald~[91] 
called the behavior of a few objects of this type a ``born again
behavior''. As a result of a ``very late'' He-shell flash in the
central star of the planetary nebula layer instead of a movement
to the white dwarf phase the star returns to the AGB stage. Recall
that a ``very late'' He-shell flash is a flash observed in a  far
evolved star, up to the state of a PN nucleus, after the final
burnout of the fuel in hydrogen shell~[89].
This recovery to the AGB is accompanied by an increase in the
luminosity of the star to $4^{\rm m}$ and a decrease of effective
temperature by tens of thousands degrees. Moreover, such  serious
changes in the parameters of the star occur over a very short time
of only about 5--15 years! Wherein, the stellar atmosphere gets
hydrogen-depleted and enriched with  the CNO-group elements, and
the $s$-process elements~[91].

In February 1996, a Japanese amateur astronomer Y.~Sakurai
registered a flash of a peculiar object, which was then named in
his honor  {\bf the Sakurai's object (V4334\,Sgr)}. During 1996 the
magnitude of the object has increased from $12\fm5$ to $11\fm2$.
Being classified by the rate of brightness variations during the
flash as a slow nova, it was referred to peculiar objects  after
its first spectra were obtained at the ESO on the 3.6-m telescope,
which did not corresponded to the expected spectrum of novae. In
March 1996, direct CCD images of the Sakurai's object taken with
the ESO 0.9-m telescope revealed a planetary nebula around the
object. Duerback and Benetti~[92], having examined the low-resolution 
spectra, concluded on a significant weakening of the neutral hydrogen 
lines at the presence of strong lines of carbon and oxygen. Considering 
the totality of the observed features they attributed the flared object 
to R\,CrB-type stars. The first high-resolution spectra of V4334\,Sgr were
obtained at the MacDonald Observatory 2.7-m telescope~[93]  and at the BTA in June 1996.
As a result of analysis of these spectra and calculating the
chemical composition by the model atmospheres~[93, 94]  a
decrease of the hydrogen abundance by 3\,dex was discovered
(thereat, it decreased by 0.7\,dex only from May to October 1996),
and a carbon excess,  and large excesses of Li and Sr, Y, Zr heavy
metals too. Kipper~[95]  thoroughly considered the variations in the 
parameters, the spectrum and chemical composition of V4334\,Sgr according 
to the results of spectroscopy of the star at the BTA after the flash. 
The detected deficiency of hydrogen and metallicity, the excess of heavy metals
and a high radial velocity indicate that the object belongs to the
old protoplanetary nebulae in the bulge of the Galaxy.

It is now assumed that the Sakurai's object has experienced a very
late He-shell flash, after which it entered into a fast final
evolutionary phase at the post-AGB stage~[89]. Therefore, the study of the
Sakurai's object gives another opportunity of testing the
theoretical modeling of the evolution of the planetary nebulae
nuclei and circumstellar envelopes, the loss of matter,
convection, stellar nucleosynthesis and  variations of the surface
chemical composition. In this regard, the Sakurai's object can be
put on a par with  FG\,Sge.

According to the  model of evolution of the star~[91], the result of an He-shell flash
should be an up to 10 times  luminosity increase and a T$_{\rm eff}$ variation from 
40\,000~K to 6300~K over 17~years. Model parameters obtained by Asplund et al.~[93] 
and Kipper and Klochkova~[94], allow us to state that V4334\,Sgr evolves faster 
(within six months its temperature dropped by 600~K). Despite the proximity of
evolutionary phases, the rate of evolution of FG\,Sge and V4334\,Sgr differs 
significantly. FG\,Sge has gone through a ``very late''  He-shell flash over 
about 100 years, and V4334\,Sgr -- 10$\div$15 times faster. Lawlor and 
MacDonald~[96]  based on the model calculations of evolution studied the phenomenon of
``very late'' He-shell flashes  and concluded that the difference in
the evolution rates of FG\,Sge and V4334\,Sgr can be explained by
assuming that FG\,Sge is observed on a slow {\it repeated} comeback to the
AGB, while V4334\,Sgr---on the fast and the first one. Lawlor and MacDonald~[96]  
emphasize that a repeated slow comeback to the AGB is possible under the condition
of a low convective mixing efficiency. A further very long
monitoring of both objects is  required to check this explanation.
Clayton et al.~[97]  believe that the ``born again'' star V605\,Aql, based on 
its spectral properties, the light curve and the evolution of the dust envelope 
is closer to V4334\,Sgr than to FG\,Sge.

In as little as a year after the flare, the formation of a dense
dust shell, which has significantly weakened the visible flux,
started around V4334\,Sgr and the object became inaccessible for
the high-resolution spectroscopy in the optical range. According
to the results of radio observations, this newly formed envelope
has a bipolar structure, in contrast to the spherical shape,
previously observed~[98].

\subsection{A star at the stage of evolution before AGB}

The list of protoplanetary nebula candidates includes a star {\bf V534\,Lyr (HD\,172324)}, 
located at a high galactic latitude ($b=18\fdg58$). In addition to a high latitude, an additional
impetus for us to study V534\,Lyr was the emission in the H\,I
lines, which was discovered  in the early work~[99]   and confirmed 
later in~[100]. This star was further studied repeatedly using different 
photometric systems and spectral methods, but so far none of the available 
publications has  any definite conclusion about its evolutionary status.

\begin{figure}[h!]
\includegraphics[scale=0.6]{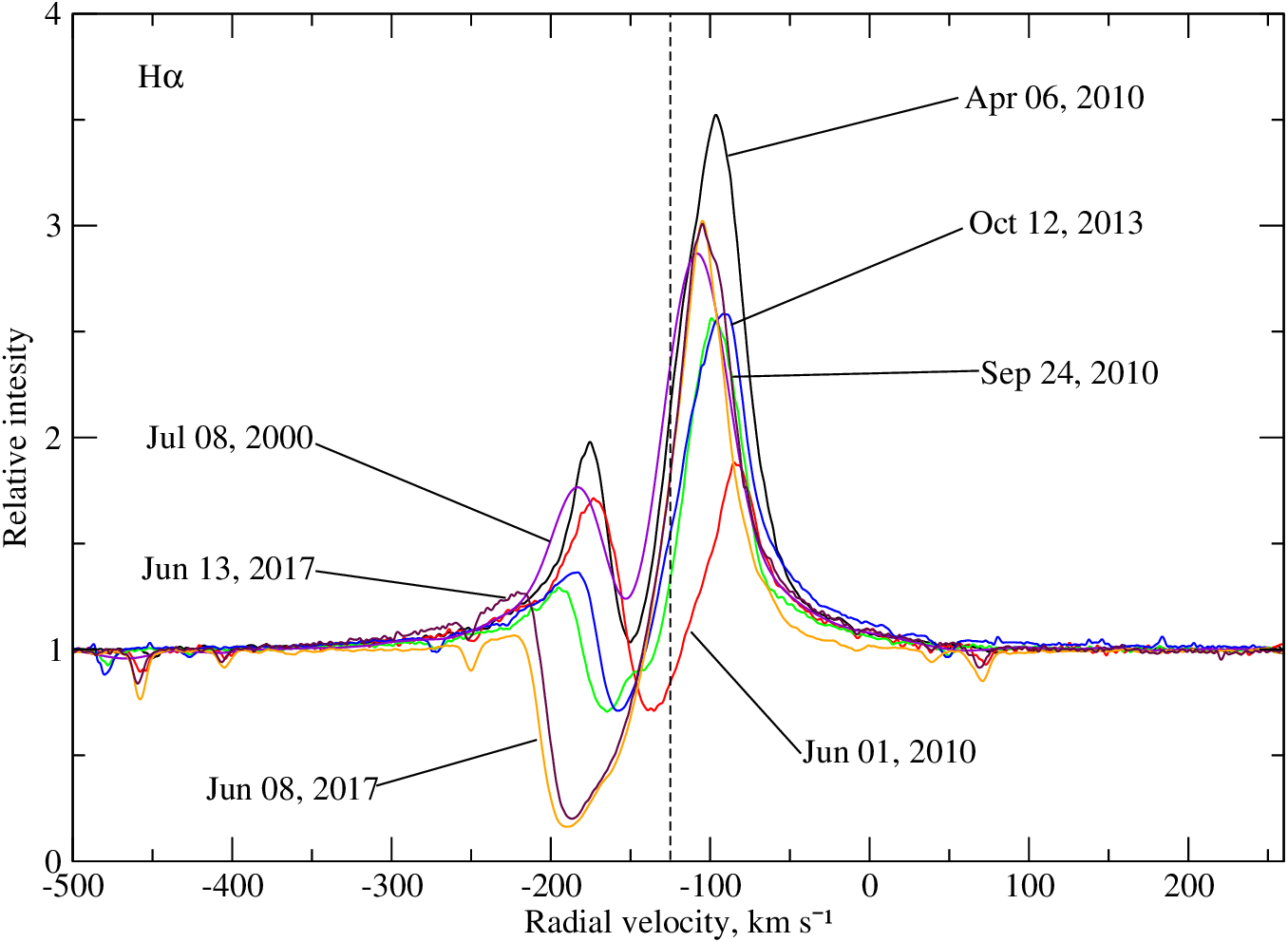}
\caption{H$\alpha$ line profiles in the V534\,Lyr spectrum for different observational sets: 
  {\it 1} --July 8, 2000; {\it 2} -- April 6, 2010;
  {\it 3} -- June 1, 2010; {\it 4} -- September 24, 2010;
  {\it 5} -- November 12, 2013; {\it 6} -- June 8, 2017;
  {\it 7} - June 13, 2017. The intermittent vertical describes the adopted systemic velocity 
  $V_{\rm sys}\approx -125$~km\,s$^{-1}$~[101]. The color version of the image
  is given in the electronic version of the article.} 
\label{V534Lyr_Halpha}
\end{figure}

Based on multiple observations at the  6-m telescope with the NES
echelle spectrograph, we thoroughly studied the features of the
optical spectrum of V534\,Lyr and the temporal behavior of the
heliocentric radial velocity corresponding to the position of all
the metal absorption components as well as the Na\,I D-lines and
the H$\alpha$ line~[101]. The analysis of the velocity field of spectral 
features of various nature revealed a low-amplitude variability of V$_r$ 
from the lines with a high excitation potential, which form in  deep layers 
of the stellar atmosphere, and allowed to estimate the systemic velocity 
V$_{\rm lsr}\approx-105$~km\,s$^{-1}$. The distance estimate of
$d\approx 6$~kpc for a high-latitude star leads to the absolute magnitude 
value of  M$_v\approx -5\fm3$, which is consistent with its spectral classification. 
A previously unknown for this star spectral phenomenon was discovered: a splitting of
profiles of selected metal absorptions at separate points of observation. In all 
cases when splitting is present in the spectrum, it reaches large values: 
$\Delta V_r=20$--$50$~km\,s$^{-1}$.

The spectral class of the star is close to A0\,Ib, and the
effective temperature T$_{\rm eff}\approx10\,500$~K. Metallicity,
reliably determined by the method of model atmospheres using the
iron group metal lines, slightly differs from the Fe abundance:
${\rm [Met/Fe]}_{\odot}=+0.06$. A nitrogen and helium excess
indicates an advanced  evolutionary stage of the star. A reduced
iron group metal abundance in combination with a high radial
velocity indicates that the star belongs to the thick disk of the
Galaxy. A set of observable features of V534\,Lyr: the presence of
probable pulsations in  deep layers of the atmosphere, a splitting
of metal absorption profiles with a low lower-level excitation
potential observed at certain times, a low metallicity, the type
and variability of the H$\alpha$ and H$\beta$ emission-absorption
profiles (see Fig.~\ref{V534Lyr_Halpha}), allows to
refer the star to variable population II stars above the HB. In
the course of subsequent evolution, it may approach the
instability  band, and then  the AGB stage. The authors~[101]  
eventually concluded on the discrepancy of the V534\,Lyr belonging 
to the post-AGB stage.

\section{CONCLUSIONS AND NEW TASKS}

The paper briefly discusses the most important issues and data
concerning the final stages of evolution of stars of different
masses and the nucleosynthesis processes during the evolution of
these objects. We presented the results obtained performing the
spectroscopic study of a sample of peculiar supergiants,
identified with galactic IR sources at the 6-m BTA telescope. The
initial list of post-AGB candidates was introduced by Kwok in the
survey~[57]. From this list we have studied all the stars reachable 
by the coordinates that have a visible brightness of  V$\le 13\fm1$. 
The main aspects of our program is a search for the evolutionary changes 
of chemical  composition of the stars that have passed the AGB stage and 
the TDU, as well as an analysis of spectral manifestations of the kinematic 
processes in their extended atmospheres and envelopes. The most significant
result of the program is the detection of an $s$-process  element
excess in seven post-AGB stars, which empirically confirms the
theory of evolution of stars of this type. In three of these stars
we have for the first time detected an outflow of  $s$-process
heavy metals into the circumstellar envelopes.

The expected lithium excess is registered in the atmospheres of peculiar 
supergiants V2324\,Cyg and V4334\,Sgr (the Sakurai's object).
The atmosphere of the ``born-again'' star V4334\,Sgr is enriched
in carbon and the $s$-process heavy metals. A  lithium excess, suspected 
for two central stars of IR sources RAFGL\,5081 and IRAS\,04296+3429,
weak in the optical range, may be a consequence of blending of
the  Li\,I\,6707 line with the Ce\,II line.

The results of  study of the kinematics of atmospheres and shells
will serve to clarifying the amount of matter produced by the
stars at the AGB and post-AGB stages and supplied to the
ISM. Up to now the models of chemical evolution of the Galaxy use
an approximation where in the spherically symmetric envelopes of
stars the matter moves only in one direction.

The results obtained during the PPN  spectroscopy  program allow
us to distinguish the priority directions in the further study of
objects at the stage of transition of a star to a planetary nebula
and related objects. First of all, a further investigation of the
`spectroscopic mimicry' problem is required, which allows peculiar
low-mass supergiants to disguise themselves as the most massive
stars of high luminosity. Secondly, we believe that we should
further focus on the spectral monitoring of selected variable and
rapidly evolving objects. Thirdly, a detailed study of the
structure of circumstellar envelopes is required, attracting the
spectropolarimetry method too to clarify the mechanisms of the
outflow and accretion of matter.

It must be emphasized that the  program of monitoring the
supergiants of various nature is  limiting for the high spectral
resolution spectroscopy even on the largest telescopes. From the
observational point of view, the task is complicated by the need
of multiple observations of variable objects, as well as high
requirements for the stability of equipment, and, in particular,
for the high accuracy of positional measurements. A new aspect of
spectroscopy of high-luminosity stars with an infrared excess,
consisting in the study of circumstellar envelopes, including
molecular spectroscopy, implies, among other things, requirements
to increase the share of high-resolution spectropolarimetry (its
first experience is already published~[102]) and provide a extremely 
high resolution $R\ge10^5$. The importance of increasing the spectral 
resolution is illustrated in Figs.~\ref{Egg_Swan} and~\ref{V5112Sgr_Ba4934}. 
Figure~\ref{Egg_Swan} clearly shows that the rotational structure of a 
molecular band can be identified only with a resolution of at least 60\,000. 
Therefore, our program stimulates the further development and sophistication 
of the spectroscopic equipment of the BTA, as well as the systems for the
reduction of spectroscopic data and  methods of its analysis.

\begin{acknowledgements}
The author thanks the Russian Foundation for Basic Research (11-02-00319\,a,  14-02-00291\,a and 
18--02--00029\,a) for the financial support. The author is grateful to the co-authors who participated 
in the implementation of the supergiant spectroscopy program at the BTA and in the preparation of joint 
publications. 
The work made use of the SIMBAD, SAO/NASA~ADS, Gaia\,DR2, AAVSO and VALD astronomical databases.
\end{acknowledgements}


\end{document}